\documentclass[aps,prc,onecolumn,preprint]{revtex4-2}
\usepackage{CJK} 
\usepackage{amsmath,amssymb} 
\usepackage{physics}
\usepackage{here} 
\usepackage[dvipdfmx]{graphicx} 
\usepackage{bm}

\begin{document}

\title{Analysis of Peierls-Yoccoz rotational energy of nuclei with semi-realistic interaction}
\author{K. Abe$^1$ and H. Nakada$^2$}
\affiliation{$^1$ Department of Physics, Graduate School of Science and Engineering, Chiba University, Yayoi-cho 1-33, Inage, Chiba 263-8522, Japan}
\affiliation{$^2$ Department of Physics, Graduate School of Science, Chiba University, Yayoi-cho 1-33, Inage, Chiba 263-8522, Japan}
\date{\today}

\begin{abstract}
The Peierls-Yoccoz (PY) rotational energy of nuclei has been analyzed by the angular-momentum projection (AMP) on the axial Hartree-Fock solutions, by using the semi-realistic effective Hamiltonian M3Y-P6.
The rotational energy is decomposed into contributions of the individual terms of the Hamiltonian, and their ratios to the total PY rotational energy are calculated.
Except for light or weakly-deformed nuclei, the ratios of the individual terms of the Hamiltonian are insensitive to nuclides and deformation.
The contributions of kinetic energies are large and close to the rigid-rotor values, although those of central forces are sizable.
For light or weakly-deformed nuclei, the ratios significantly depend on nuclei and deformation.
The contributions of noncentral forces are not negligible.
Regardless of nuclides, the attractive forces decrease the moment-of-inertia, and the repulsive forces increase it.
A general formula for the PY rotational energy is derived, which suggests that higher-order terms of the cumulant expansion play roles in the rotational energy and the moment-of-inertia for light or weakly-deformed nuclei.
\end{abstract}

\maketitle

\section{Introduction}

The rotational band is a well-known energy spectrum, $E_{\mathrm{x}}(J)\approx J(J+1) / 2\,\mathcal{I}$ \cite{BM98}.
It is observed experimentally over a wide range of the nuclear chart, including not only stable nuclei but also unstable ones \cite{NNDC}.
It indicates that the intrinsic state of nuclei is deformed and rotates with the moment-of-inertia $\mathcal{I}$.

From a microscopic standpoint, nuclei have been described self-consistently by the mean-field (MF) theory, such as the Hartree-Fock (HF) and the Hartree-Fock-Bogolyubov (HFB) approximations \cite{RS80}.
Because nuclei are isolated systems, the nuclear Hamiltonian has rotational symmetry, and the angular momentum is a good quantum number in energy eigenstates.
However, spontaneous breaking of the rotational symmetry often occurs in the MF approximation.
The rotational symmetry breaking of the MF state corresponds with a deformation of the intrinsic state.
The deformed intrinsic state in nuclei is not observed directly.
The Nambu-Goldstone (NG) mode is accompanied by the symmetry breaking, and it restores the corresponding symmetry in energy eigenstates.
The restoration of the rotational symmetry corresponds with a whole rotational motion of the deformed nuclei.

Several methods that treat the rotation of nuclei have been developed.
As a microscopic theory, the cranking model \cite{RS80} has been proposed.
The Inglis formula \cite{In5455} and the Belyaev formula \cite{Be59} have been derived for the moment-of-inertia from the cranking model.
The Thouless-Valatin formula \cite{TV62} has been obtained in connection to the random phase approximation (RPA).
The angular-momentum projection (AMP) has been developed \cite{RS80,PY57,Yo57,PT62,Ve63,Ve64,OY66,Ka68,BB69,RER02,BB21,SDRRY21}, in which the degenerate intrinsic states along the NG mode are superposed.
The $J(J+1)$ rule of the excitation energy with the moment-of-inertia is derived from the AMP under a reasonable approximation for well-deformed heavy nuclei \cite{PY57,Yo57,Ve63,Ve64,Ka68,RS80}.
However, for light or weakly-deformed nuclei, it is not sufficiently clear whether the same arguments hold.
It should also be noted that the rotation significantly affects the intrinsic state, as handled in the cranking model \cite{PT62,Ka68,RS80} and the variation-after-projection (VAP) schemes \cite{RS80}.

In the classical mechanics, the rotational energy arises from the kinetic energy.
The rotational energy of nuclei should be formed from the effective Hamiltonian including the nucleonic interaction.
In principle, the nucleonic interaction originates from the quantum chromodynamics (QCD) \cite{PS95}.
However, it is not yet easy to derive the nucleonic interaction from the QCD which is applicable to a variety of nuclei with good accuracy.
Because the nucleonic interactions are effectively mediated by mesons, they are represented by the Yukawa functions \cite{Yu35}.
The Michigan-three-range-Yukawa (M3Y)-type interactions \cite{BBML77,Na03,Na13,Na20} are composed of the Yukawa functions except for density-dependent terms.
The parameters of the effective interactions have been determined based on the G-matrix with some phenomenological modifications \cite{Na13}.
In this respect, the M3Y-type interactions are {\it semi-realistic} effective interactions.
The tensor force is also included, whose parameters are fixed from the G-matrix.
It has been pointed out via the nuclear matter response functions that the M3Y-type interactions are free from most of the instabilities, which sometimes occur in other interactions \cite{DPN21}.
The MF calculations using the M3Y-P6 have described well the magic number of nuclei over a wide range of the nuclear chart \cite{NS1416}.
Because it has a certain connection to the bare nucleonic interaction and is applicable to self-consistent MF calculations including deformation, the M3Y-type interaction is suitable for analyzing the rotational energy of nuclei.

Under these backgrounds, we shall re-examine from the microscopic point of view how the rotational energy of nuclei is formed.
The AMP is applied to the MF wave functions obtained by self-consistent axial-HF calculations using the effective interaction M3Y-P6.
In the present study, the energies produced from a fixed HF intrinsic state are inspected.
Namely, we restrict ourselves to the energies arising solely from the rotation of the HF intrinsic state, separating them out from the effects of the rotation on the intrinsic state and ignoring the pair correlations.
As obtained by the AMP of Peierls and Yoccoz \cite{PY57}, we call this energy {\it Peierls-Yoccoz (PY) rotational energy}.
It should be kept in mind that the PY rotational energy is not enough to describe the rotational spectra in actual nuclei \cite{PT62}.
The contributions of the individual terms of the semi-realistic Hamiltonian to the PY rotational energy are focused on; in concrete, those of the kinetic energy, the density-independent and dependent central forces, the LS force, the tensor force, and the central part of the one-pion-exchange-potential (OPEP), which is the longest-range term and an example of spin-dependent channels.
It is noted that the nucleonic interactions have spin-dependence, both in these central and noncentral channels, which could contribute to the rotational energy.
Additionally, we present a general formulation for the PY rotational energy. 
Compared with the previous formulas \cite{PY57,Yo57,Ve63,Ve64,Ka68,RS80}, we find additional terms which could be important for light or weakly-deformed nuclei.

\section{Theoretical background}

\subsection{Theory of AMP and rotation}
\label{subsec:AMP_rot}

The AMP is the method by which an intrinsic state is projected on angular-momentum eigenstates \cite{RS80}.
In the following, we assume that $\hat{\mathcal{S}}$ is a rotational scalar, and the intrinsic state is an eigenstate of $\hat{J}_z$ whose eigenvalue is $M$.
The intrinsic state $\ket{ \Phi_M }$ is expanded by angular-momentum eigenstates $\ket{JM}$, where we omit indices other than $J$ and $M$ for simplicity,
\begin{equation}\label{eq:expansion_JM}
\ket*{ \Phi_M } 
= 
\sum_J \ket*{ JM } \braket*{ JM }{ \Phi_M }.
\end{equation}
The Wigner (small) d-function \cite{VMK,JJ94,RS80} is defined by matrix elements of a rotational operator around the $y$-axis with the angle $\beta$,
\begin{equation}\label{eq:dfunc}
d^{(J)}_{MK}(\beta)
:= 
\mel*{ J M } { e^{-i \hat{J}_y \beta} } { J K }.
\end{equation}
On the standard phase convention of the angular-momentum, $d^{(J)}_{MK}(\beta)$ takes a real number.
The expectation values of the scalar operator $\hat{\mathcal{S}}$ on the angular-momentum eigenstates are obtained as follows \cite{RS80},
\begin{equation}\label{eq:SJ}
\ev*{ \hat{\mathcal{S}} }{ J }
=\,
\frac{
\displaystyle \int_0^\pi d\beta 
\sin \beta \,
d^{(J)}_{M M} \left( \beta \right)
\ev*{ \hat{\mathcal{S}} \, e^{-i \hat{J}_y \beta} }{ \Phi_M }
}
{
\displaystyle \int_0^\pi d\beta 
\sin \beta \, 
d^{(J)}_{M M} \left( \beta \right)
\ev*{ e^{-i \hat{J}_y \beta} }{ \Phi_M }
},
\end{equation}
where we omit the index $M$ on the LHS of Eq. \eqref{eq:SJ}.

By using the property \cite{VMK} $d^{(J)}_{M M}(-\beta)=d^{(J)}_{M M}(\beta)$,
the following relation is derived for $\hat{\mathcal{S}}$ and $\ket{\Phi_M}$,
\begin{equation}\label{eq:even_property}
\begin{split}
\ev*{ \hat{\mathcal{S}} \, e^{-i \hat{J}_y \beta} }{ \Phi_M }
=&
\sum_{J}
| \braket{ JM }{ \Phi_M } |^2
\ev*{ \hat{\mathcal{S}} }{ J }
d^{(J)}_{MM}(\beta)\\
=&
\ev*{ \hat{\mathcal{S}} \, e^{i \hat{J}_y \beta} }{ \Phi_M }.
\end{split}
\end{equation}
Therefore, $\ev*{ \hat{\mathcal{S}} \, e^{-i \hat{J}_y \beta} }{ \Phi_M }$ is an even function of $\beta$, and we have $\ev*{ \hat{\mathcal{S}} \, \hat{J}_y^{\,2n+1} }{ \Phi_M }
= 0$ for $n=0,1,2,...$, whose particular case is $\ev*{ \hat{J}_y }{ \Phi_M } = 0$.

The following function is defined,
\begin{equation}\label{eq:def_S01}
\mathcal{S}^{01}(\beta)
:=
\frac{
\ev*{ \hat{\mathcal{S}} \, e^{-i \hat{J}_y \beta} }{ \Phi_M }
}
{
\ev*{ e^{-i \hat{J}_y \beta} }{ \Phi_M }
},
\end{equation}
which is also an even function of $\beta$.
The correlation function between operators $\hat{A}$ and $\hat{B}$ is defined as $C[\hat{A},\hat{B}]:=\ev*{ \hat{A} \hat{B} }-\ev*{ \hat{A} } \ev*{ \hat{B} }$, where the bracket $\ev*{\,}$ represents the expectation value at $\ket{\Phi_M}$.
The above $\mathcal{S}^{01}(\beta)$ is related to the correlation function between $\hat{\mathcal{S}}$ and $\hat{J}^{\,2}_y$,
\begin{equation}\label{eq:C}
\begin{split}
\left. \frac{d^{\,2}}{d\beta^{\,2}}\,
\mathcal{S}^{01}(\beta) \right| _{\beta=0}
=\,
-\,C[ \hat{\mathcal{S}}, \hat{J}_y^{\,2} ].
\end{split}
\end{equation}
The fluctuation of an operator $\hat{A}$ is defined as $\sigma [ \hat{A} ]:=\sqrt{ C[\hat{A},\hat{A}] }$.
The overlap function $\ev*{ e^{-i \hat{J}_y \beta} }{ \Phi_M }$ is related to the fluctuation of $\hat{J}_y$ as,
\begin{equation}\label{eq:sigma}
\begin{split}
\left. \frac{d^{\,2}}{d\beta^{\,2}}\,
\ev*{ e^{-i \hat{J}_y \beta} }{ \Phi_M } \, \right| _{\beta=0}
=
-\, ( \sigma [ \hat{J}_y ] )^2.
\end{split}
\end{equation}
Concerning $C[ \hat{\mathcal{S}}, \hat{J}_y^{\,2} ]$ and $( \sigma [ \hat{J}_y ] )^2$, we have
\begin{subequations}\label{eq:J_y^2_J^2}
\begin{align}
&C[ \hat{\mathcal{S}}, \hat{J}_x^{\,2} ]
=
C[ \hat{\mathcal{S}}, \hat{J}_y^{\,2} ]
=
\frac{1}{2} \, C[ \hat{\mathcal{S}}, \hat{\bm{J}}^{\,2} ],
\label{eq:C_J_y^2_J^2}\\
&( \sigma [ \hat{J}_x ] )^2
=
( \sigma [ \hat{J}_y ] )^2
=
\frac{1}{2} \, \left( \ev*{ \hat{\bm{J}}^{\,2} }{ \Phi_M } - M^2 \right).
\label{eq:sigma_J_y^2_J^2}
\end{align}
\end{subequations}

Let us restrict ourselves to the even-even nuclei with $M=0$.
Extension to the $M \ne 0$ case is almost straightforward.
We further assume that the state $\ket{\Phi_0}$ has the following symmetry,
\begin{equation}\label{eq:def_Rsym}
\hat{\mathcal{R}} \ket*{\Phi_0} = \ket*{\Phi_0},~~~
\mathcal{\hat{R}} := e^{-i\hat{J}_y \pi},
\end{equation}
then $\braket*{ J0 }{ \Phi_0 }=0$ for odd $J$.
By using Eq. \eqref{eq:even_property}, the following equation is derived,
\begin{equation}\label{eq:pi-beta}
\ev*{ \mathcal{\hat{S}} \, e^{-i \hat{J}_y (\pi-\beta)} }{ \Phi_0 }
=
\ev*{ \mathcal{\hat{S}} \, e^{-i \hat{J}_y \beta} }{ \Phi_0 }.
\end{equation}
From Eq. \eqref{eq:pi-beta} and $d^{(J)}_{00}(\pi-\beta)
=(-)^{J}d^{(J)}_{00}(\beta)$, the range of $\beta$ integration in Eq. \eqref{eq:SJ} can be reduced to $[0,\pi/2]$ \cite{Ve63,RER02}.

The $J(J+1)$ rule of rotational energy and the moment-of-inertia connected with Eq. \eqref{eq:SJ} were discussed in Refs. \cite{PY57,Yo57,Ve63,Ve64,Ka68,RS80}.
If the overlap function $\ev*{ e^{-i\hat{J}_y \beta} }{ \Phi_0 }$ has a sharp peak at $\beta \approx 0$, the energy spectrum is close to the $J(J+1)$ rule.
However, it is not always clear whether $\ev*{ e^{-i\hat{J}_y \beta} }{ \Phi_0 }$ has a sharp peak at $\beta \approx 0$.
In the following, we present a more general argument on the rotational energy than those in Refs. \cite{PY57,Ve63} by using the cumulant expansion \cite{Ku62}.
This formulation is useful in some cases, as will be discussed in Sec. \ref{sec:Numerical_results}.

We expand $d^{(J)}_{00}(\beta)$ by the power series of $\beta$,
\begin{subequations}\label{eq:Taylor_dfun_c2n}
\begin{align}
&d^{(J)}_{00}(\beta)
=
\sum_{n=0}^{\infty} c_{2n} \beta^{2n},
\label{eq:Taylor_dfun}\\
&c_{2n}
=\,
\frac{ (-)^n }{ (2n)! } \ev*{ \hat{J}_y^{\,2n} }{ J0 }
=\,
\frac{
\ev*{ (\hat{J}_{+}-\hat{J}_{-})^{2n} }{ J0 }
}
{
(2n)! \, 2^{2n}
}.
\label{eq:c2n}
\end{align}
\end{subequations}
Equation \eqref{eq:c2n} leads to,
\begin{subequations}
\begin{align}
c_0 =& 1,
\label{eq:c0}\\
c_2 =& -\frac{1}{2! \, 2} \, J(J+1),
\label{eq:c2}\\
c_4 =& \frac{1}{4! \, 2^3} \, J(J+1) \left[ 3J(J+1)-2\right ], \cdots.
\label{eq:c4}
\end{align}
\end{subequations}
The coefficient $c_{2n}$ depends only on $J$.
The cumulant of operators $\hat{X}_1$, $\hat{X}_2$, $\cdots$, $\hat{X}_n$ \cite{Ku62} is defined by,
\begin{equation}\label{eq:def_cumulant}
\ev*{ \hat{X}_1 ; \cdots ; \hat{X}_n }_{\mathrm{cum}}
:=
\frac{\partial}{\partial t_1}
\cdots
\frac{\partial}{\partial t_n}
\left.
\ln
\ev{ \exp \left(\sum_{i=1}^{n} t_i {\hat{X}}_i \right) }
\right|_{t_1=\cdots=t_n=0},
\end{equation}
where $[\hat{X}_i,\hat{X}_j]=0$ for all $i$ and $j$.
The following equation is derived,
\begin{equation}\label{eq:XetY}
\begin{split}
\frac{ \ev*{\hat{X}_1 e^{t\hat{X}_2}} }{ \ev*{e^{t\hat{X}_2}} }
=&
\left.
\sum_{n=0}^{\infty}
\frac{t^n}{n!}
\frac{\partial}{\partial t_1}
\frac{\partial^n}{\partial t_2^{\,n}}
\ln \ev*{ e^{t_1\hat{X}_1 + t_2\hat{X}_2} }
\right|_{t_1=t_2=0}\\
=&
\sum_{n=0}^{\infty}
\frac{t^n}{n!}
\ev*{ \hat{X}_1;\underbrace{ \hat{X}_2 ; \cdots ; \hat{X}_2 }_{n} }_{\mathrm{cum}}.
\end{split}
\end{equation}
Via Eq. \eqref{eq:XetY}, $\mathcal{S}^{01}(\beta)$ in Eq. \eqref{eq:def_S01} is expanded as follows,
\begin{subequations}\label{eq:S01_s2n}
\begin{align}
&\mathcal{S}^{01}(\beta)
=
\sum_{n=0}^{\infty} s_{2n} \beta^{2n},
\label{eq:Taylor_S01}\\
&s_{2n}
=\,
\frac{ (-)^n } { (2n)! }
\ev*{ \hat{\mathcal{S}} ; \underbrace{ \hat{J}_y ; \cdots ; \hat{J}_y }_{2n} }{ \Phi_0 }_{\mathrm{cum}}.
\label{eq:s2n_cumulant}
\end{align}
\end{subequations}
Equation \eqref{eq:s2n_cumulant} leads to,
\begin{subequations}
\begin{align}
s_0 =& \ev*{ \hat{\mathcal{S}} }{ \Phi_0 },
\label{eq:s0}\\
s_2 =& -\frac{1}{2!} \, C[ \hat{\mathcal{S}}, \hat{J}_y^{\,2} ],
\label{eq:s2}\\
s_4 =& \,\frac{1}{4!}
\left( C[ \hat{\mathcal{S}}, \hat{J}_y^{\,4} ]
-6 \, C[ \hat{\mathcal{S}}, \hat{J}_y^{\,2} ]
(\sigma [ \hat{J}_y ])^2
\right), \cdots.
\label{eq:s4}
\end{align}
\end{subequations}
The coefficient $s_{2n}$ is independent of $J$, depending only on $\ket*{\Phi_0}$ and $\hat{\mathcal{S}}$.
By defining the following quantities,
\begin{subequations}\label{eq:N_2n_Lambda_2n}
\begin{align}
N_{2n}
&:=
\int_0^{\pi/2}d\beta \sin \beta \, \beta^{2n}
\ev*{ e^{-i \hat{J}_y \beta} }{ \Phi_0 },
\label{eq:N_2n}\\
\varLambda_{2n}
&:=
\frac{N_{2n}}{N_0},
~~~(n=0,1,2,\cdots),
\label{eq:Lambda_2n}
\end{align}
\end{subequations}
which are determined only by $\ket{\Phi_0}$, Eq. \eqref{eq:SJ} is rewritten as follows,
\begin{equation}\label{eq:JSJ_Lambda}
\begin{split}
\ev*{ \hat{\mathcal{S}} }{ J }
=&\,
\frac{
\displaystyle \int_0^{\pi/2} d\beta \sin \beta \,
d^{(J)}_{00} \left( \beta \right)
\ev*{ e^{-i \hat{J}_y \beta} }{ \Phi_0 }
\mathcal{S}^{01}(\beta)
}
{
\displaystyle \int_0^{\pi/2} d\beta \sin \beta \,
d^{(J)}_{00} \left( \beta \right)
\ev*{ e^{-i \hat{J}_y \beta} }{ \Phi_0 }
}\\
=&\,
\frac{ \displaystyle \sum_{m,n=0}^{\infty}
c_{2m} s_{2n} N_{2m+2n} }
{ \displaystyle \sum_{\ell=0}^{\infty} c_{2\ell} N_{2\ell} } \\
=&\,
\frac{
\displaystyle \sum_{m,n=0}^{\infty} c_{2m} s_{2n} \varLambda_{2m+2n}
}
{
\displaystyle \sum_{\ell=0}^{\infty} c_{2\ell} \varLambda_{2\ell}
}.\\
\end{split}
\end{equation}
For $J=0$, $c_{2n}$ vanishes for $n\ge1$, and the following equation is obtained,
\begin{equation}\label{eq:0S0}
\ev*{ \hat{\mathcal{S}} }{ 0 }
=
\sum_{n=0}^{\infty}
s_{2n} \varLambda_{2n}
=
\ev*{ \hat{\mathcal{S}} }{ \Phi_0 }
+
\sum_{n=1}^{\infty}
s_{2n} \varLambda_{2n}.
\end{equation}
In the present expression, the energy difference $\ev*{ \hat{H} }{ J } - \ev*{ \hat{H} }{ 0 }$ for an axial-HF state $\ket{\Phi_0}$, where $\hat{H}$ is the Hamiltonian, is the PY rotational energy.

For the denominator on the RHS of Eq. \eqref{eq:JSJ_Lambda}, we have
\begin{equation}\label{eq:epsilon0}
\sum_{n=0}^{\infty} c_{2n} \varLambda_{2n}
=
1+\sum_{n=1}^{\infty} c_{2n} \varLambda_{2n}
=
\frac{1} {2J+1}
\left|
\frac{ \braket*{ J0 }{ \Phi_0 } } { \braket*{ 00 }{ \Phi_0 } }
\right|^2.
\end{equation}
The inequality $\left| \sum_{n=1}^{\infty} c_{2n} \varLambda_{2n} \right| < 1$ is usually satisfied.
We have numerically confirmed via the RHS of Eq. \eqref{eq:epsilon0} that this inequality is indeed satisfied in all cases handled in Sec. \ref{sec:Numerical_results}.
We then expand Eq. (\ref{eq:JSJ_Lambda}) as follows,
\begin{equation}\label{eq:SJN_expand1}
\begin{split}
\ev*{ \hat{\mathcal{S}} }{ J }
=&
\left(
\sum_{m,n=0}^{\infty}
s_{2n} c_{2m} \varLambda_{2n+2m}
\right)
\left(
1 + \sum_{\ell=1}^{\infty} c_{2\ell} \varLambda_{2\ell}
\right)^{-1} \\
=&
\left(
\sum_{m=0}^{\infty} \sum_{n=0}^{\infty}
s_{2n} c_{2m} \varLambda_{2n+2m}
\right)\\
&\times
\left[
1 -
\sum_{\ell=1}^{\infty} c_{2\ell} \varLambda_{2\ell}
+
\left(
\sum_{\ell=1}^{\infty} c_{2\ell} \varLambda_{2\ell}
\right)^2
-\cdots
\right].
\end{split}
\end{equation}
In order to analyze $J$-dependence of $\ev*{ \hat{\mathcal{S}} } { J }$, with taking account of $c_{2n} \sim J^{2n}$, it is appropriate to arrange Eq. \eqref{eq:SJN_expand1} by $c_{2n}$ as follows,
\begin{equation}\label{eq:SJN_expand2}
\begin{split}
\ev*{ \hat{\mathcal{S}} }{ J }
=&
\left(
\sum_{n=0}^{\infty}
s_{2n} \varLambda_{2n}
+ c_2 \sum_{n=0}^{\infty}
s_{2n} \varLambda_{2n+2}
+ c_4 \sum_{n=0}^{\infty}
s_{2n} \varLambda_{2n+4}
+\cdots
\right)
\\
&\times \left[
1 - ( c_2 \varLambda_2 + c_4 \varLambda_4 + \cdots )
+
( c_2 \varLambda_2 )^2
+\cdots
\right] \\
=&
\sum_{n=0}^{\infty}
s_{2n} \varLambda_{2n}
+
c_2 \sum_{n=1}^{\infty} s_{2n}
\left(
\varLambda_{2n+2}-\varLambda_{2n} \varLambda_2
\right)\\
&+
c_4
\left(
\sum_{n=1}^{\infty} s_{2n}
\left(
\varLambda_{2n+4} - \varLambda_{2n} \varLambda_{4}
\right)
\right)\\
&-\left( c_2 \right)^2
\left(
\sum_{n=1}^{\infty} s_{2n}
\left[
\varLambda_{2n+2} \varLambda_{2} - \varLambda_{2n} (\varLambda_{2})^2
\right]
\right)
+\cdots.
\end{split}
\end{equation}
Compared with the Kamlah expansion \cite{Ka68,RS80}, the cumulant expansion of Eq. \eqref{eq:S01_s2n} enables a more organized expansion of the rotational energy.
We call the $n \ge 2$ terms of $c_{2n}$ in Eq. \eqref{eq:SJN_expand2} {\it higher}-$\mathit{c_{2n}}$-{\it terms}, and those of $s_{2n}$ {\it higher}-$\mathit{s_{2n}}$-{\it terms}.
If the higher-$c_{2n}$-terms are neglected, Eq. \eqref{eq:SJN_expand2} is approximated as,
\begin{equation}\label{eq:J(J+1)}
\begin{split}
\ev*{ \hat{\mathcal{S}} }{ J }
\approx
\ev*{ \hat{\mathcal{S}} }{ 0 }
+
\frac{J(J+1)}{2\, \mathcal{I}[\mathcal{S}]},
\end{split}
\end{equation}
where,
\begin{equation}\label{eq:def_moi}
\frac{1}{\mathcal{I}[\hat{\mathcal{S}}]}
:=
\sum_{n=1}^{\infty}
s_{2n}
\left[
-
\frac{1}{2} (\varLambda_{2n+2}-\varLambda_{2n} \varLambda_2)
\right].
\end{equation}
For $\mathcal{\hat{S}}=\hat{H}$, the parameter $\mathcal{I}[\hat{H}]$ is interpreted as the moment-of-inertia.
If the higher-$s_{2n}$-terms are neglected in Eqs. \eqref{eq:0S0} and \eqref{eq:def_moi}, $\ev*{ \hat{\mathcal{S}} }{ 0 }$ and $\mathcal{I}[\hat{\mathcal{S}}]$ are approximated as,
\begin{subequations}\label{eq:0S0_moi_approx0}
\begin{align}
\ev*{ \hat{\mathcal{S}} }{ 0 }
\approx&
\ev*{ \hat{\mathcal{S}} }{ \Phi_0 }
-
\frac{1}{2}\,
C[ \hat{\mathcal{S}}, \hat{J}_y^{\,2} ]
\varLambda_2,
\label{eq:0S0_approx0}\\
\frac{1}{\mathcal{I}[\hat{\mathcal{S}}]}
\approx&\,
\frac{1}{4}\,
C[ \hat{\mathcal{S}}, \hat{J}_y^{\,2} ] \,
[\varLambda_{4}- (\varLambda_2)^2].
\label{eq:moi_approx0}
\end{align}
\end{subequations}
Equation \eqref{eq:moi_approx0} gives the moment-of-inertia of Peierls and Yoccoz \cite{PY57,Ve63}.
Equation \eqref{eq:def_moi} is regarded as a generalization of Eq. \eqref{eq:moi_approx0}.
From Eqs. \eqref{eq:0S0}, \eqref{eq:def_moi}, and \eqref{eq:0S0_moi_approx0}, it is noticed that the higher-$s_{2n}$-terms may contribute to $\ev*{ \hat{\mathcal{S}} }{ 0 }$ and $\mathcal{I}[\hat{\mathcal{S}}]$.
We shall see such a situation in Sec. \ref{subsec:influence}.
Further approximation based on the Gaussian approximation \cite{Yo57,Ve64,Ka68,RS80} with the higher-$s_{2n}$-terms is discussed in Appendix \ref{ap:Gauss_approx}.

\subsection{Semi-realistic effective Hamiltonian}
\label{subsec:effective_H}

We have implemented AMP calculations in Eq. \eqref{eq:SJ} for the axial-HF solutions using the semi-realistic interaction M3Y-P6 \cite{Na13,Na20,SNM16,MN18}.
It is the first application of the M3Y-type interactions to the AMP calculations.

Because nuclei are finite and isolate systems, their effective Hamiltonian should have rotational, parity, and time-reversal symmetry, with number conservation.
We assume that the individual terms of the Hamiltonian also have isospin symmetries except for the Coulomb force.
The Hamiltonian is composed of the following terms,
\begin{equation}\label{eq:Hamiltonian}
H=K + V_{\mathrm{nucl}} + V_{\mathrm{Coulomb}} - H_{\mathrm{c.m.}}.
\end{equation}
The kinetic energy is $K=\sum_i \bold{p}_i^2/2M$, the nucleonic interaction between two nucleons is $V_{\mathrm{nucl}}=\sum_{i<j} v_{ij}$, the Coulomb interaction between protons is denoted as $V_{\mathrm{Coulomb}}$, and the center-of-mass term is $H_{\mathrm{c.m.}}=\bold{P}^2 /2AM$, with the total momentum $\bold{P}=\sum_i \bold{p}_i$ and the mass number $A=Z+N$.
The effective nucleonic interaction is formed by the following terms,
\begin{subequations}\label{eq:nuclear_forces}
\begin{align}
V_{\mathrm{nucl}}
&=
V^{(\mathrm{C})} +
V^{(\mathrm{LS})} +
V^{(\mathrm{TN})} +
V^{(\mathrm{C\rho})},
\label{eq:int_total}\\
V^{(\mathrm{X})}
&=
\sum_{i<j} v_{ij}^{(\mathrm{X})},~~~
(\mathrm{X}=\mathrm{C, LS, TN, C\rho}),
\label{eq:int_total_def}
\end{align}
\end{subequations}
where $V^{(\mathrm{C})}$, $V^{(\mathrm{LS})}$, and $V^{(\mathrm{TN})}$ are the central, the LS, and the tensor forces.
The central density-dependent term is distinguished from $V^{(\mathrm{C})}$ and represented by $V^{(\mathrm{C\rho})}$.
The individual terms of Eq. \eqref{eq:int_total} have the following forms,
\begin{widetext}
\begin{equation}\label{eq:interaction}
\begin{split}
v_{ij}^{(\mathrm{C})}
=&
\sum_{n} \left(
t^{\mathrm{(SE)}}_{n} P_{\mathrm{SE}}
+
t^{\mathrm{(TE)}}_{n} P_{\mathrm{TE}}
+
t^{\mathrm{(SO)}}_{n} P_{\mathrm{SO}}
+
t^{\mathrm{(TO)}}_{n} P_{\mathrm{TO}}
\right)
f^{\mathrm{(C)}}_{n}(r_{ij}), \\
v_{ij}^{(\mathrm{LS})}
=&
\sum_{n} \left(
t^{\mathrm{(LSE)}}_{n} P_{\mathrm{TE}}
+
t^{\mathrm{(LSO)}}_{n} P_{\mathrm{TO}}
\right)
f^{\mathrm{(LS)}}_{n}(r_{ij})\,
\bold{L}_{ij} \cdot (\bold{s}_i + \bold{s}_j), \\
v_{ij}^{(\mathrm{TN})}
=&
\sum_{n} \left(
t^{\mathrm{(TNE)}}_{n} P_{\mathrm{TE}}
+
t^{\mathrm{(TNO)}}_{n} P_{\mathrm{TO}}
\right)
f^{\mathrm{(TN)}}_{n}(r_{ij})\,
r^2_{ij} \, S_{ij}, \\
v_{ij}^{(\mathrm{C\rho})}
=&
\left(
t^{\mathrm{(SE)}}_{\rho} P_{\mathrm{SE}} \cdot \left[ \rho(\bold{r}_i) \right]^{\alpha^{(\mathrm{SE})}}
+
t^{\mathrm{(TE)}}_{\rho} P_{\mathrm{TE}} \cdot \left[ \rho(\bold{r}_i) \right]^{\alpha^{(\mathrm{TE})}}
\right)
\delta(\bold{r}_{ij}),
\end{split}
\end{equation}
\end{widetext}
where
$ \bold{r}_{ij} := \bold{r}_{i} - \bold{r}_{j} $,
$ r_{ij} := |\bold{r}_{ij}| $,
$ \bold{\hat{r}}_{ij} := \bold{r}_{ij} / r_{ij} $,
$ \bold{p}_{ij} := (\bold{p}_{i} - \bold{p}_{j})/2 $,
$ \bold{L}_{ij} := \bold{r}_{ij} \times \bold{p}_{ij} $,
and
\begin{equation}
\begin{split}
S_{ij}
&:=
3(\boldsymbol{\sigma}_i \cdot \bold{\hat{r}}_{ij}) 
(\boldsymbol{\sigma}_j \cdot \bold{\hat{r}}_{ij})
-
\boldsymbol{{\sigma}}_i \cdot \boldsymbol{\sigma}_j. \\
\end{split}
\end{equation}
The spin- and isospin-exchange operators between two nucleons are defined as,
\begin{equation}
P_{\sigma}
:=
\frac{1+\boldsymbol{{\sigma}}_i \cdot \boldsymbol{\sigma}_j}{2},
~~~~~
P_{\tau}
:=
\frac{1+\boldsymbol{{\tau}}_i \cdot \boldsymbol{\tau}_j}{2},
\end{equation}
then the projection operators on singlet-even (SE), triplet-even (TE), singlet-odd (SO), and triplet-odd (TO) channels are defined as,
\begin{equation}
\begin{split}
P_{\mathrm{SE}} &:= \frac{1-P_{\sigma}}{2} \frac{1+P_{\tau}}{2},
~~~~~
P_{\mathrm{TE}} := \frac{1+P_{\sigma}}{2} \frac{1-P_{\tau}}{2},\\
P_{\mathrm{SO}} &:= \frac{1-P_{\sigma}}{2} \frac{1-P_{\tau}}{2},
~~~~~
P_{\mathrm{TO}}: = \frac{1+P_{\sigma}}{2} \frac{1+P_{\tau}}{2}.
\end{split}
\end{equation}
We use the Yukawa function $f_{n}(r) =\mathrm{e}^{-\mu_{n}r}/(\mu_{n}r)$ for the radial functions, except for $v_{ij}^{(\mathrm{C\rho})}$.
The longest-range term in $v^{\mathrm{(C)}}_{ij}$ is fixed to be that of the OPEP.
This central OPEP, which is denoted by $V^{\mathrm{(OPEP)}}$, is an example of spin-dependent interactions.
The values of the parameters for M3Y-P6 are given in Ref. \cite{Na13}.

\subsection{Implementation of AMP}

In this work, we apply the projection-after-variation for the AMP to the axial-HF solutions.
The numerical method of Eq. \eqref{eq:SJ} has been discussed in Refs. \cite{RS80,RER02}.
The intrinsic state could gradually change for increasing $J$, often accompanied by breaking of the axial and the time-reversal symmetry.
While these effects can be handled in the cranking model \cite{PT62,Ka68,RS80} and in the VAP schemes \cite{RS80}, they are ignored in the present study, and we focus on rotational energy arising from a fixed intrinsic state in this paper, as stated in Introduction.
Furthermore, the AMP calculations for the HFB solutions are left for future works.

The Gaussian expansion method (GEM) has been applied in which complex-range Gaussian bases are used to expand the radial part of the single-particle wave function \cite{NS02,Na08}.
The angular function is the spinor-spherical-harmonics.
The advantages of the GEM in the MF calculations are taken over to the AMP calculations.
Additionally, the spherical bases enable precise numerical calculations of the AMP relatively easily.
Some details of the AMP calculations for non-orthogonal bases are given in Appendix \ref{ap:AMP}.

The parity and the time-reversal operators are represented as $\hat{\mathcal{P}}$ and $\hat{\mathcal{T}}$, respectively.
For the sake of simplicity, we say ``$\hat{\mathcal{O}}$ symmetry'' when $\hat{\mathcal{O}} \ket*{\Phi_0} = \ket*{\Phi_0}$ is satisfied.
In the MF calculations, $\hat{\mathcal{P}}$, $\hat{\mathcal{T}}$, $\hat{\mathcal{R}}$ (see Eq. \eqref{eq:def_Rsym}), and axial symmetries are assumed.
Owing to the $\hat{\mathcal{R}}\hat{\mathcal{T}}$ symmetry, the elements of the matrices $\mathsf{U}$ and $\mathsf{V}$ in Eq. \eqref{eq:transUV} which represent the MF solutions are real numbers.
The MF state $\ket{ \Phi_0 }$ is a direct product of the parts having specific isospin and parity,
\begin{equation}\label{eq:phi_tz_pi}
\ket*{ \Phi_0 } =
\ket*{ \Phi_0 (p+) } \otimes
\ket*{ \Phi_0 (p-) } \otimes
\ket*{ \Phi_0 (n+) } \otimes
\ket*{ \Phi_0 (n-) }.
\end{equation}

The overlap function $\ev*{ e^{ -i \hat{J}_y \beta} }{ \Phi_0 }$ has been calculated by the Onishi formula \cite{OY66,BB69,RS80} (see Eq. \eqref{eq:onishi}).
The sign problem of the Onishi formula is well-known, and some solutions have been proposed \cite{NW83,Ro09}.
In the present cases, the sign problem does not occur owing to non-negativity of the overlap function, as proven in Appendix \ref{ap:nonneg}.

There is also a problem in the density-dependent coefficients in $v_{ij}^{\mathrm{(C\rho)}}$ in the AMP calculations \cite{BH08,BB21}.
The density-dependent term in Eq. \eqref{eq:interaction} is not a rotational scalar when the density does not have the spherical symmetry.
In the present calculations, the standard treatment in Ref. \cite{RER02} has been adopted, replacing the density $\rho(\bold{r})$ in Eq. \eqref{eq:interaction} with ``{\it generalized density}\,'' $\bar{\rho}(\bold{r};\beta)$ which is defined as
\begin{subequations}
\begin{align}
\bar{\rho}(\bold{r} ; \beta)
&:=
\sum_{\tau} \sum_{\sigma}
\rho^{01}(\bold{r} \sigma \tau ; \beta),
\label{eq:generalized_dens}\\
\rho^{01}(\bold{r} \sigma \tau ; \beta)
&:=
\sum_{k_1 k_2}
\rho^{01}_{k_1 k_2} (\beta)
\varphi_{k_1} (\bold{r} \sigma \tau)
\varphi_{k_2}^{\ast} (\bold{r} \sigma \tau),
\end{align}
\end{subequations}
where $\rho^{01}_{k_1 k_2}(\beta)$ is the generalized density matrix in Eq. \eqref{eq:defGDM01}, and $\varphi_k(\bold{r} \sigma \tau) := \braket*{ \bold{r} \sigma \tau }{ k }$, $\sigma$ and $\tau$ are the spin and the isospin indices.
According to the $\hat{\mathcal{T}}$ symmetry, the generalized density in Eq. \eqref{eq:generalized_dens} becomes a real number.
However, there is a case in which $\bar{\rho}(\bold{r};\beta)$ becomes negative and its fractional power $\bar{\rho}^{\,\alpha}(\bold{r};\beta)$ may become multivalued.
In the M3Y-P6 interaction, the fractional powers $\alpha$ are $\alpha^{\mathrm{(SE)}}=1$ and $\alpha^{\mathrm{(TE)}}=1/3$ \cite{Na13}.
The phase of $\bar{\rho}^{\,\alpha}(\bold{r};\beta)$ has been chosen negative when $\bar{\rho}(\bold{r};\beta)$ is negative.
For a rotational scalar $\hat{\mathcal{S}}$, the following equation should hold,
\begin{equation}\label{eq:Jsum}
\ev*{ \hat{\mathcal{S}} }{ \Phi_0 }
=
\sum_J (2J+1)
\displaystyle \int_0^{\pi/2} d\beta
\sin \beta
\, d^{(J)}_{00} \left( \beta \right)
\ev*{ \hat{\mathcal{S}} \, e^{-i \hat{J}_y \beta} }{ \Phi_0 }.
\end{equation}
There is no mathematical guarantee that Eq. \eqref{eq:Jsum} is fulfilled for $\hat{V}^{\mathrm{(C\rho)}}$ when the LHS is evaluated by the MF state and the RHS is calculated with $\bar{\rho}(\bold{r};\beta)$.
Nonetheless, in the present calculations, Eq. \eqref{eq:Jsum} is satisfied for $\hat{V}^{\mathrm{(C\rho)}}$ comparably well to those for the other terms of the Hamiltonian in Eqs. \eqref{eq:Hamiltonian} and \eqref{eq:nuclear_forces}.

\section{Numerical results}

\label{sec:Numerical_results}

In the present work, the AMP calculations in Eq. \eqref{eq:SJ} have been applied to deformed $_{12}$Mg \cite{SNM16,Na20}, $_{40}$Zr \cite{MN18}, $_{60}$Nd and $_{62}$Sm isotopes, including stable and unstable nuclei.
An important evidence for the deformation is their ratios of excitation energies $E_{\mathrm{x}}(4^+)/E_{\mathrm{x}}(2^+)$ close to $10/3$ \cite{NNDC,DS13,SY11,PC17}.
$^{24}_{12}$Mg is known as a light stable well-deformed nucleus.
$^{34-38}_{~~~~12}$Mg are well-deformed unstable nuclei \cite{DS13}.
$^{40}_{12}$Mg is near the neutron dripline \cite{BA07}, and a deformed halo structure of the intrinsic state has been suggested \cite{NT18}.
$^{80}_{40}$Zr is a deformed unstable nucleus near the proton dripline \cite{LC87}.
$^{100-110}_{~~~~~~\,40}$Zr are neutron-rich well-deformed nuclei \cite{NNDC,SY11,PC17}.
$^{150-154}_{~~~~~~\,60}$Nd and $^{152-156}_{~~~~~~\,62}$Sm are well-known as deformed nuclei \cite{NNDC,BM98,St16}.

\subsection{Contribution of individual terms of effective Hamiltonian to PY rotational energy}
\label{subsec:contribution}

In this subsection, we present composition of the PY rotational energy.
The expectation values of the individual terms of the effective Hamiltonian at angular-momentum eigenstates are calculated.
The following quantity is defined via Eq. \eqref{eq:SJ},
\begin{equation}\label{eq:S_x}
\mathcal{S}_{\mathrm{x}}(J^+)
:=
\ev*{ \hat{\mathcal{S}} }{ J^+ }
-
\ev*{ \hat{\mathcal{S}} }{ 0^+ }.
\end{equation}
For $\mathcal{\hat{S}}=\hat{H}$, where $\hat{H}$ is the Hamiltonian, $\mathcal{S}_{\mathrm{x}}(J^+)$ corresponds with the excitation energy (\textit{i.e.}, the PY rotational energy),
\begin{equation}\label{eq:E_x}
E_{\mathrm{x}}(J^+)
=
\ev*{ \hat{H} }{ J^+ }
-
\ev*{ \hat{H} }{ 0^+ }.
\end{equation}
By taking $\hat{\mathcal{S}}$ to be individual terms of $\hat{H}$, the values $\mathcal{S}_{\mathrm{x}}(J^+)$ give their contributions to the rotational energy.
In the following, $\hat{\mathcal{S}}$ is an element of the following set,
\begin{equation}\label{eq:S_in}
\hat{\mathcal{S}} \in \{ \hat{H},\hat{K},\hat{V}^{\mathrm{(C)}},\hat{V}^{\mathrm{(LS)}},\hat{V}^{\mathrm{(TN)}},\hat{V}^{\mathrm{(C\rho)}},\hat{V}^{\mathrm{(OPEP)}} \},
\end{equation}
where each ingredient has been defined in Sec. \ref{subsec:effective_H}.

\begin{figure}[H]
\centering
\includegraphics[width=8.6cm,clip]{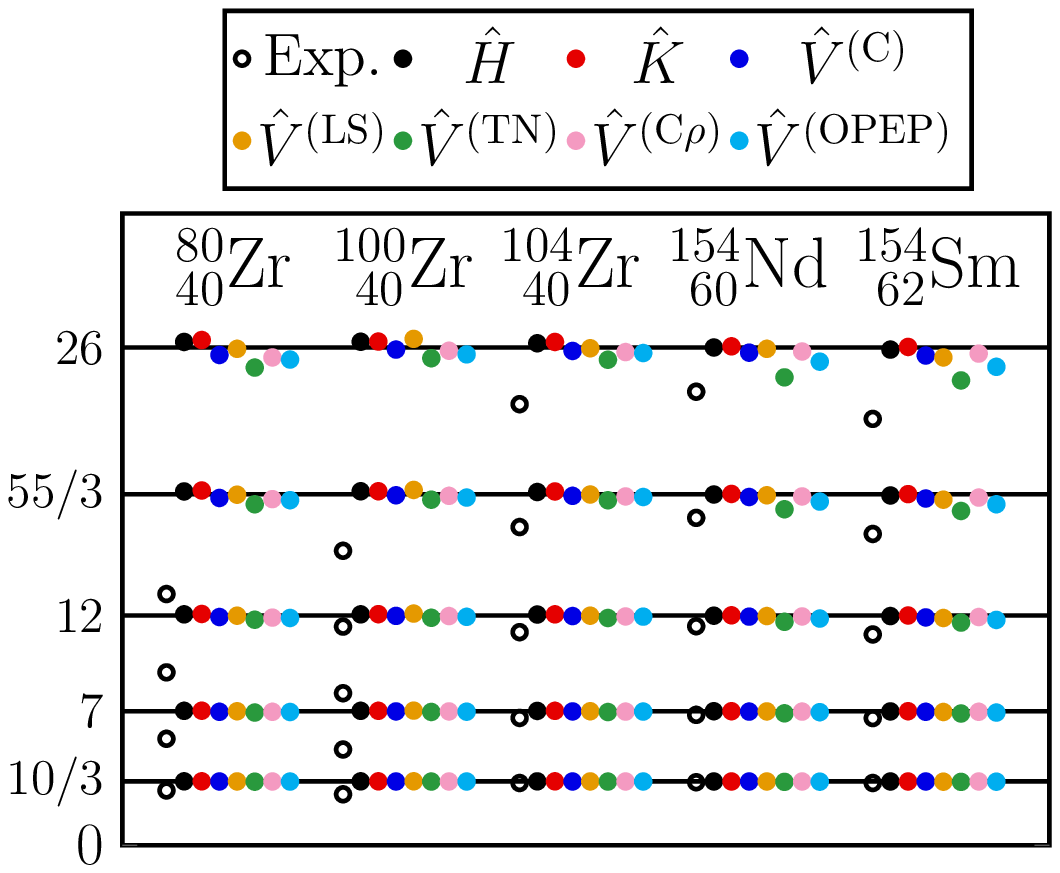}
\caption[]
{
The ratios $E_{\mathrm{x}}(J^+)/E_{\mathrm{x}}(2^+)$ and $\mathcal{S}_{\mathrm{x}}(J^+)/\mathcal{S}_{\mathrm{x}}(2^+)$ for the deformed $^{80,100,104}_{~~~~~~~~40}$Zr, $^{154}_{~60}$Nd, and $^{154}_{~62}$Sm nuclei at their lowest minima.
The open circles are the ratios $E_{\mathrm{x}}(J^+)/E_{\mathrm{x}}(2^+)$ of experiments \cite{NNDC}, and the circles filled in black are those obtained by the present work.
The ratios $\mathcal{S}_{\mathrm{x}}(J^+)/\mathcal{S}_{\mathrm{x}}(2^+)$ are also shown for
$\hat{\mathcal{S}}=\hat{K}$ (red circles),
$\hat{V}^{\mathrm{(C)}}$ (blue circles),
$\hat{V}^{\mathrm{(LS)}}$ (yellow circles),
$\hat{V}^{\mathrm{(TN)}}$ (green circles),
$\hat{V}^{\mathrm{(C\rho)}}$ (pink circles), and
$\hat{V}^{\mathrm{(OPEP)}}$ (sky-blue circles).
The rigid-rotor value of $J(J+1)/6$ is presented by the horizontal lines.
}
\label{ratio_ZrNdSm}
\end{figure}

\begin{figure}[H]
\centering
\includegraphics[width=8.6cm,clip]{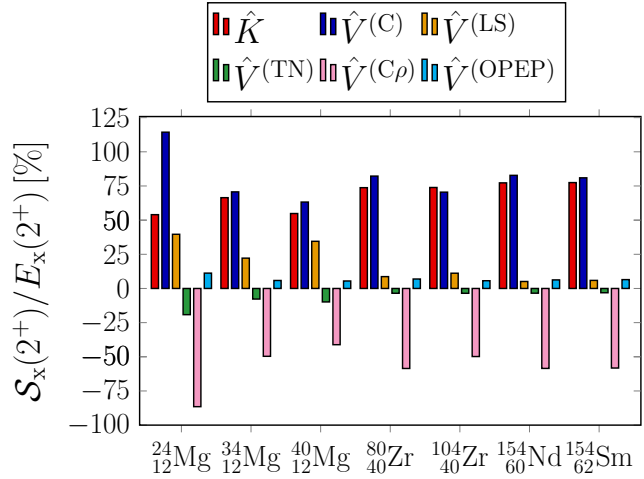}
\caption[]
{
The ratios $\mathcal{S}_{\mathrm{x}}(2^+) / E_{\mathrm{x}}(2^+)$ for the deformed nuclei at their lowest minima.
}
\label{percentage_HF_pro}
\end{figure}

The ratios of the individual terms in the effective Hamiltonian $\mathcal{S}_{\mathrm{x}}(J^+)/\mathcal{S}_{\mathrm{x}}(2^+)$ have been calculated as well as the ratio of the total excitation energy $E_{\mathrm{x}}(J^+)/E_{\mathrm{x}}(2^+)$.
The results for the deformed $_{40}$Zr, $^{154}_{~60}$Nd, and $^{154}_{~62}$Sm nuclei at their lowest minima are shown in Fig. \ref{ratio_ZrNdSm}.
The ratios $E_{\mathrm{x}}(4^+)/E_{\mathrm{x}}(2^+)$ obtained by the present work are close to those of the experiments and $10/3$.
The values of $\mathcal{S}_{\mathrm{x}}(J^+)$ are negative for $\hat{V}^{\mathrm{(TN)}}$ and $\hat{V}^{\mathrm{(C\rho)}}$ as will be shown in Fig. \ref{percentage_HF_pro}.
As well as
$E_{\mathrm{x}}(J^+)/E_{\mathrm{x}}(2^+)$, $\mathcal{S}_{\mathrm{x}}(J^+)/\mathcal{S}_{\mathrm{x}}(2^+)$ well obeys the $J(J+1)$ rule, up to high angular-momentum $J\approx12$.
The experimental values gradually get smaller than the $J(J+1)$ line in many nuclei at high $J$.
Additional quantum correlations such as gradual change of the intrinsic state should be considered in order to reproduce experimental values more accurately.

\begin{figure}[H]
\centering
\includegraphics[width=8.6cm,clip]{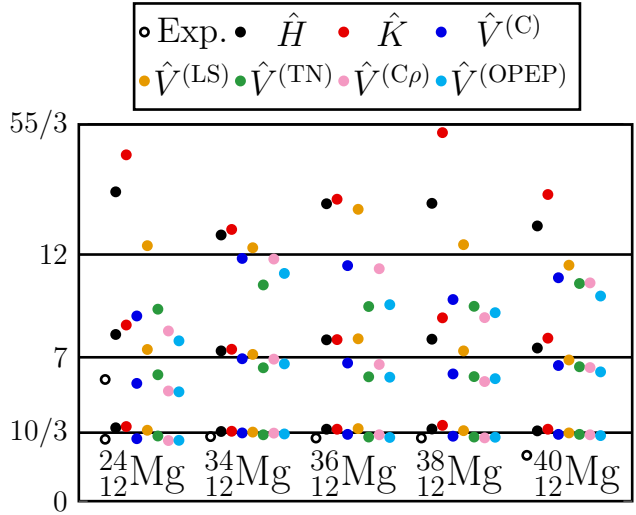}
\caption[]
{
The ratios $E_{\mathrm{x}}(J^+)/E_{\mathrm{x}}(2^+)$ and $\mathcal{S}_{\mathrm{x}}(J^+)/\mathcal{S}_{\mathrm{x}}(2^+)$ for the deformed $_{12}$Mg isotopes at their lowest minima.
See Fig. \ref{ratio_ZrNdSm} for conventions.
The experimental values of $E_{\mathrm{x}}(J^+)/E_{\mathrm{x}}(2^+)$ are taken from Refs. \cite{NNDC,DS13,CF19}.
For $_{12}^{40}$Mg, the spin and parity of the excited states have not been confirmed yet.
}
\label{ratio_Mg}
\end{figure}

In Fig. \ref{ratio_Mg}, the ratios $E_{\mathrm{x}}(J^+)/E_{\mathrm{x}}(2^+)$ and $\mathcal{S}_{\mathrm{x}}(J^+)/\mathcal{S}_{\mathrm{x}}(2^+)$ for the deformed $_{12}$Mg isotopes at their lowest minima are shown.
Note that for $^{40}_{12}$Mg, the spin and parity of the excited states have not yet been confirmed experimentally.
The ratios $E_{\mathrm{x}}(4^+)/E_{\mathrm{x}}(2^+)$ obtained by the present work are close to those of the experiments and $10/3$ except for $^{40}_{12}$Mg.
The ratios $\mathcal{S}_{\mathrm{x}}(4^+) / \mathcal{S}_{\mathrm{x}}(2^+)$ are also close to $10/3$, as well.
As $J$ increases, the ratios deviate from the $J(J+1)$ lines.

Because the ratios $\mathcal{S}_{\mathrm{x}}(4^+) / \mathcal{S}_{\mathrm{x}}(2^+)$ are close to $10/3$ irrespective of $\hat{\mathcal{S}}$ and nuclides in Figs. \ref{ratio_ZrNdSm} and \ref{ratio_Mg}, we focus on compositions of the first excitation energies $E_{\mathrm{x}}(2^+)$.
Figure \ref{percentage_HF_pro} shows the ratios $\mathcal{S}_{\mathrm{x}}(2^+) / E_{\mathrm{x}}(2^+)$ at their lowest minima, all of which have prolate shapes.
Except for the $_{12}$Mg region, these ratios are insensitive to nuclides.
The contributions of $\hat{K}$, $\hat{V}^{\mathrm{(C)}}$, and $\hat{V}^{\mathrm{(C\rho)}}$ are about $75$\%, $75$\%, and $-50$\%, respectively.
The large positive contribution of $\hat{K}$ is harmonious with the rotational energy in classical mechanics.
Both $\hat{V}^{\mathrm{(C)}}$ and $\hat{V}^{\mathrm{(C\rho)}}$ give sizable contributions, although they tend to cancel to a certain extent.
The contributions of $\hat{V}^{\mathrm{(LS)}}$ and $\hat{V}^{\mathrm{(TN)}}$ are small.
These noncentral forces mainly contribute near the surface of nuclei.
Therefore, these forces become relatively small compared to the central forces when the mass number increases.
In the $_{12}$Mg region, the ratios significantly depend on nuclei.
The LS force widens the rotational band, and the tensor force narrows it, whose ratios are large compared to those of $^{80,104}_{~~~~40}$Zr, $^{154}_{~60}$Nd, and $^{154}_{~62}$Sm nuclei.
Regardless of nuclides, $\hat{V}^{\mathrm{(C)}}$ and $\hat{V}^{\mathrm{(LS)}}$ act attractively, and $\hat{V}^{\mathrm{(TN)}}$ and $\hat{V}^{\mathrm{(C\rho)}}$ do repulsively on the binding energies.
The contributions of the former are positive, and those of the latter are negative for the rotational energies.
In other words, the attractive forces decrease the moment-of-inertia of nuclei, and the repulsive forces increase it.
The contributions of $\hat{V}^{\mathrm{(OPEP)}}$ are about 10 \% at most.
The contributions of $\hat{V}_{\mathrm{Coulomb}}$ and $\hat{H}_{\mathrm{c.m.}}$ to the excitation energies are no more than a few percent.
The latter results indicate that the center-of-mass motion and the rotational motion, both of which are NG modes in the MF approximation, hardly couple each other.

\begin{figure}[H]
\centering
\includegraphics[width=8.6cm,clip]{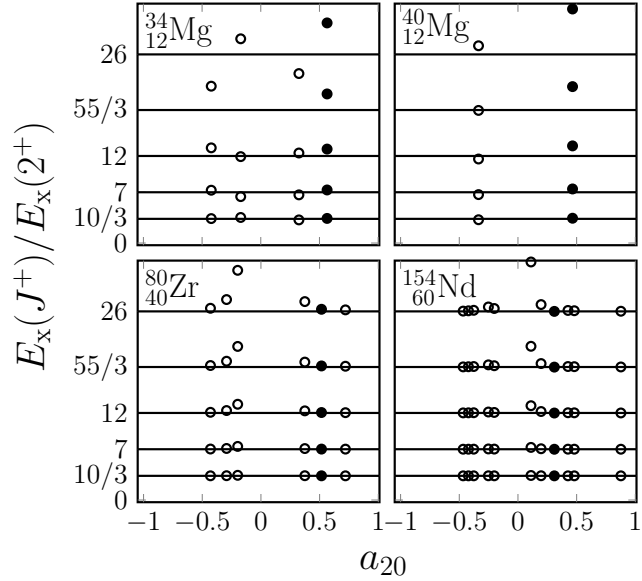}
\caption[]
{
The $a_{20}$ (quadrupole deformation parameter) dependence of $E_{\mathrm{x}}(J^+)/E_{\mathrm{x}}(2^+)$ for the $^{34}_{12}$Mg, $^{40}_{12}$Mg, $^{80}_{40}$Zr, and $^{154}_{~60}$Nd nuclei.
Their lowest minima are represented by the filled circles.
The $J(J+1)/6$ values are presented by the horizontal lines.
}
\label{ratio_def_para}
\end{figure}

We define the quadrupole deformation parameter $a_{20}$ as follows \cite{BM98},
\begin{equation}\label{eq:def_para}
a_{20}:=\frac{q_0}{1.09 A^{5/3}},
\end{equation}
where $q_0$ is the mass quadrupole moment of the MF state in units of $\mathrm{fm}^2$ \cite{SNM16}.
Figure \ref{ratio_def_para} shows the $a_{20}$ dependence of $E_{\mathrm{x}}(J^+)/E_{\mathrm{x}}(2^+)$ for axial-HF solutions of the $^{34}_{12}$Mg, $^{40}_{12}$Mg, $^{80}_{40}$Zr, and $^{154}_{~60}$Nd nuclei, including their local minima.
For low $J$, the ratios $E_{\mathrm{x}}(J^+)/E_{\mathrm{x}}(2^+)$ are close to those given by the $J(J+1)$ rule, which indicates that the approximation in Eq. \eqref{eq:J(J+1)} is good.
For the $^{80}_{40}$Zr and $^{154}_{~60}$Nd nuclei, the ratios become closer to the $J(J+1)$ line up to high $J$ as $|a_{20}|$ increases.
For the $^{34,40}_{~~~12}$Mg nuclei or the minima having small $|a_{20}|$ values, the ratios get deviating from the $J(J+1)$ line as $J$ increases, though the intrinsic states are fixed.
This deviation indicates that the higher-$c_{2n}$-terms are not negligible in Eq. \eqref{eq:SJN_expand2}.

\begin{widetext}
\begin{figure}[H]
\centering
\includegraphics[width=17.8cm,clip]{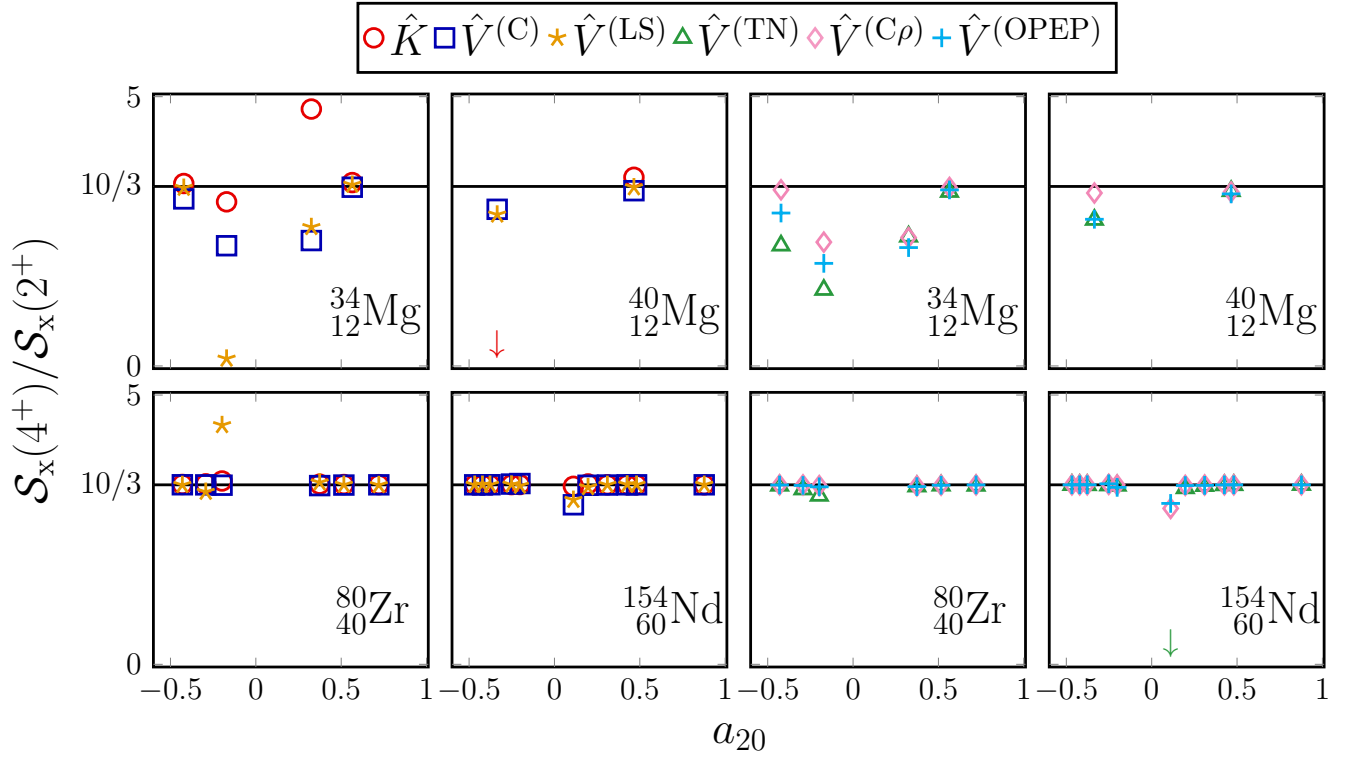}
\caption[]
{
The $a_{20}$ dependence of $\mathcal{S}_{\mathrm{x}}(4^+)/\mathcal{S}_{\mathrm{x}}(2^+)$ for
$\hat{\mathcal{S}}=\hat{K}$ (red circle),
$\hat{V}^{\mathrm{(C)}}$ (blue squares),
$\hat{V}^{\mathrm{(LS)}}$ (yellow stars),
$\hat{V}^{\mathrm{(TN)}}$ (green triangles),
$\hat{V}^{\mathrm{(C\rho)}}$ (pink diamonds), and
$\hat{V}^{\mathrm{(OPEP)}}$ (sky-blue pluses).
}
\label{ratio_divide_def_para2}
\end{figure}
\end{widetext}

In Fig. \ref{ratio_divide_def_para2}, the $a_{20}$ dependence of $\mathcal{S}_{\mathrm{x}}(4^+)/\mathcal{S}_{\mathrm{x}}(2^+)$ is shown.
For $^{80}_{40}$Zr and $^{154}_{~60}$Nd, the ratios are close to $10/3$, which is almost independent of $a_{20}$ and $\hat{\mathcal{S}}$ with only a few exceptions.
For $a_{20}=0.56$ of $^{34}_{12}$Mg and $a_{20}=0.47$ of $^{40}_{12}$Mg, the ratios are also close to $10/3$, which is independent of $\hat{\mathcal{S}}$.
However, at the other minima of $^{34,40}_{~~~12}$Mg, the results strongly depend on $a_{20}$ and $\hat{\mathcal{S}}$.

\begin{figure}[H]
\centering
\includegraphics[width=8.6cm,clip]{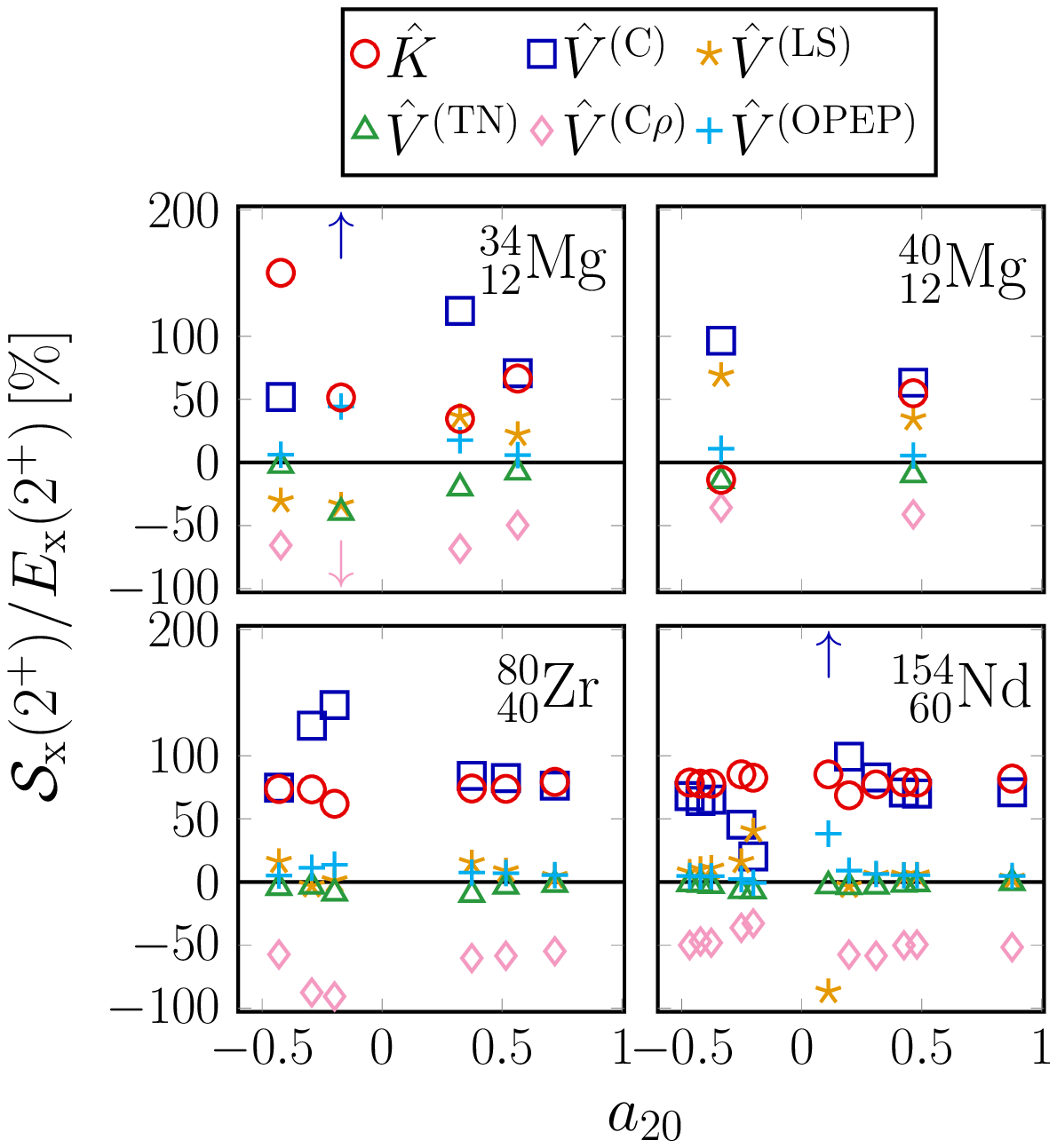}
\caption[]
{
The $a_{20}$ dependence of $\mathcal{S}_{\mathrm{x}}(2^+) / E_{\mathrm{x}}(2^+)$.
}
\label{percentage_def_para}
\end{figure}

In Fig. \ref{percentage_def_para}, the $a_{20}$ dependence of the ratio $\mathcal{S}_{\mathrm{x}}(2^+) / E_{\mathrm{x}}(2^+)$ is shown.
For $^{80}_{40}$Zr and $^{154}_{~60}$Nd, the ratios become almost constant for $a_{20}$.
In particular, the ratios of $\hat{K}$ are almost independent of the deformation parameter.
The contributions of $\hat{V}^{\mathrm{(C)}}$ and $\hat{V}^{\mathrm{(LS)}}$ become positive, and those of $\hat{V}^{\mathrm{(C\rho)}}$ and $\hat{V}^{\mathrm{(TN)}}$ do negative apart from a few exceptions.
The ratios of these interactions fluctuate in the regions where $|a_{20}|$ is not large.
For $^{34,40}_{~~~12}$Mg nuclei, the ratios strongly depend on the individual MF states.
For $^{40}_{12}$Mg, we find an extraordinary result that the ratio of $\hat{K}$ is negative at $a_{20}=-0.34$.
At this MF state, $J=2$ gives the lowest value of $\mathcal{S}(J^+)$ for $\hat{\mathcal{S}}=\hat{K}$.

\begin{figure}[H]
\centering
\includegraphics[width=8.6cm,clip]{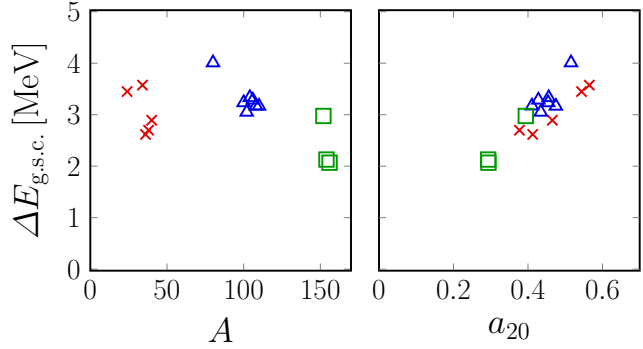}
\caption[]
{
The ground-state correlations $\varDelta E_{\mathrm{g.s.c.}}$ obtained by the AMP calculations for the deformed $^{24,34-40}_{~~~~~~~\,12}$Mg (red crosses), $^{80,100-110}_{~~~~~~~~~\,40}$Zr (blue triangles), and $^{152-156}_{~~~~~~\,62}$Sm (green squares) isotopes at their lowest minima.
The horizontal axes are the mass number $A$ and the quadrupole deformation parameter $a_{20}$.
}
\label{GScorr_A_q0}
\end{figure}

We define the ground-state correlation as follows (see Eq. \eqref{eq:0S0}),
\begin{equation}\label{eq:g.s.c.}
\varDelta E_{\mathrm{g.s.c.}} := \ev*{ \hat{H} }{ \Phi_0 } - \ev*{ \hat{H} }{ 0^+ }.
\end{equation}
The values of $\varDelta E_{\mathrm{g.s.c.}}$ obtained by the AMP calculations for the deformed $_{12}$Mg, $_{40}$Zr, and $_{62}$Sm isotopes at their lowest minima are shown in Fig. \ref{GScorr_A_q0}.
While $\varDelta E_{\mathrm{g.s.c.}}$ is not sensitive to the mass number, it correlates well to $a_{20}$ with the correlation coefficient $0.89$.
Thus, $\varDelta E_{\mathrm{g.s.c.}}$ increases as deformation of nuclei does, as expected.

\subsection{Influence of higher-$c_{2n}$-terms and higher-$s_{2n}$-terms}
\label{subsec:influence}

In this subsection, we investigate influence of heigher-$c_{2n}$-terms and higher-$s_{2n}$-terms in Eq. \eqref{eq:SJN_expand2}, for the $^{34}_{12}$Mg, $^{40}_{12}$Mg, $^{80}_{40}$Zr, and $^{154}_{~60}$Nd nuclei, including their local minima.

\begin{widetext}
\begin{figure}[H]
\centering
\includegraphics[width=17.8cm,clip]{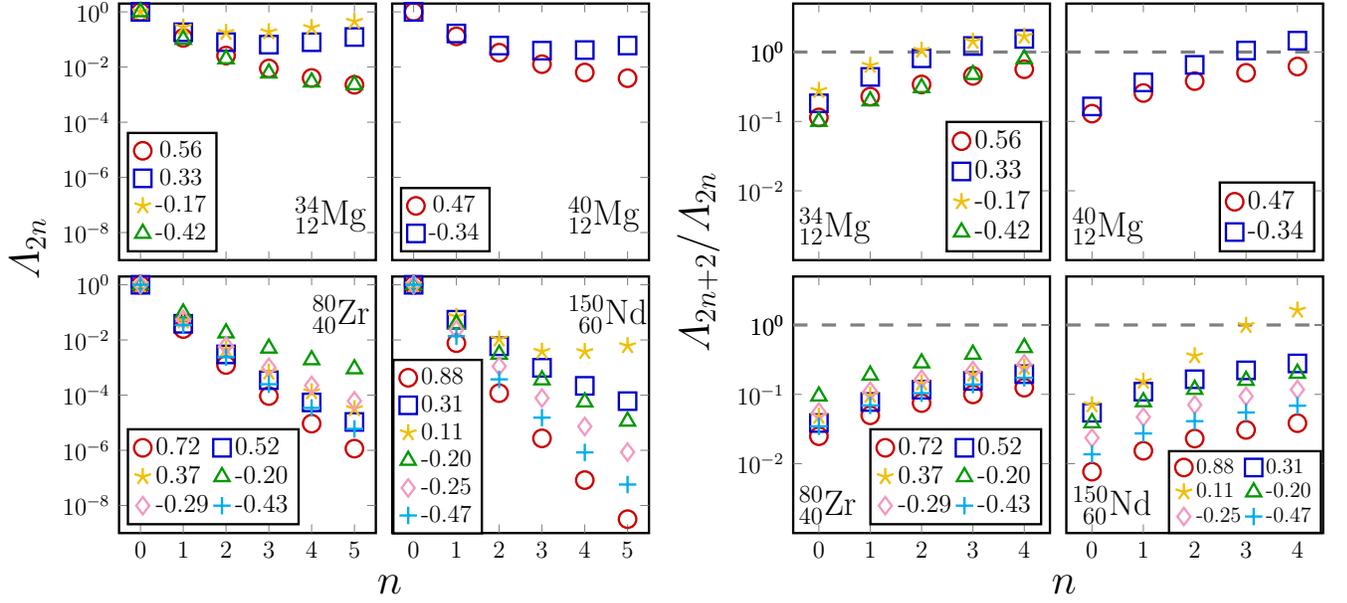}
\caption[]
{
The $\varLambda_{2n}$ and $\varLambda_{2n+2}/\varLambda_{2n}$ values at the various $a_{20}$ values shown in the insets.
}
\label{Lambda_2n}
\end{figure}
\end{widetext}

It is important for the $J(J+1)$ rule in Eq. \eqref{eq:J(J+1)} that the higher-$c_{2n}$-terms are small compared to the $c_2$ term in Eq. \eqref{eq:SJN_expand2}.
To examine influence of the higher-$c_{2n}$-terms and the higher-$s_{2n}$-terms in Eq. \eqref{eq:SJN_expand2}, the $\varLambda_{2n}$ and $\varLambda_{2n+2}/\varLambda_{2n}$ values are shown in Fig. \ref{Lambda_2n}.
For the well-deformed minima of $^{80}_{40}$Zr and $^{154}_{~60}$Nd, the values of $\varLambda_{2n}$ are small compared to the $^{34,40}_{~~~12}$Mg nuclei or the weakly-deformed minima.
As $|a_{20}|$ and the mass number increases, $\varLambda_{2n+2}/\varLambda_{2n}$ decreases for fixed $n$.
Small $\varLambda_{2n+2}/\varLambda_{2n}$ values help both the $J(J+1)$ rule and the approximation of Eq. \eqref{eq:0S0_moi_approx0}, although $c_{2n}$ and $s_{2n}$ also play roles.

\begin{widetext}
\begin{figure}[H]
\centering
\includegraphics[width=17.3cm,clip]{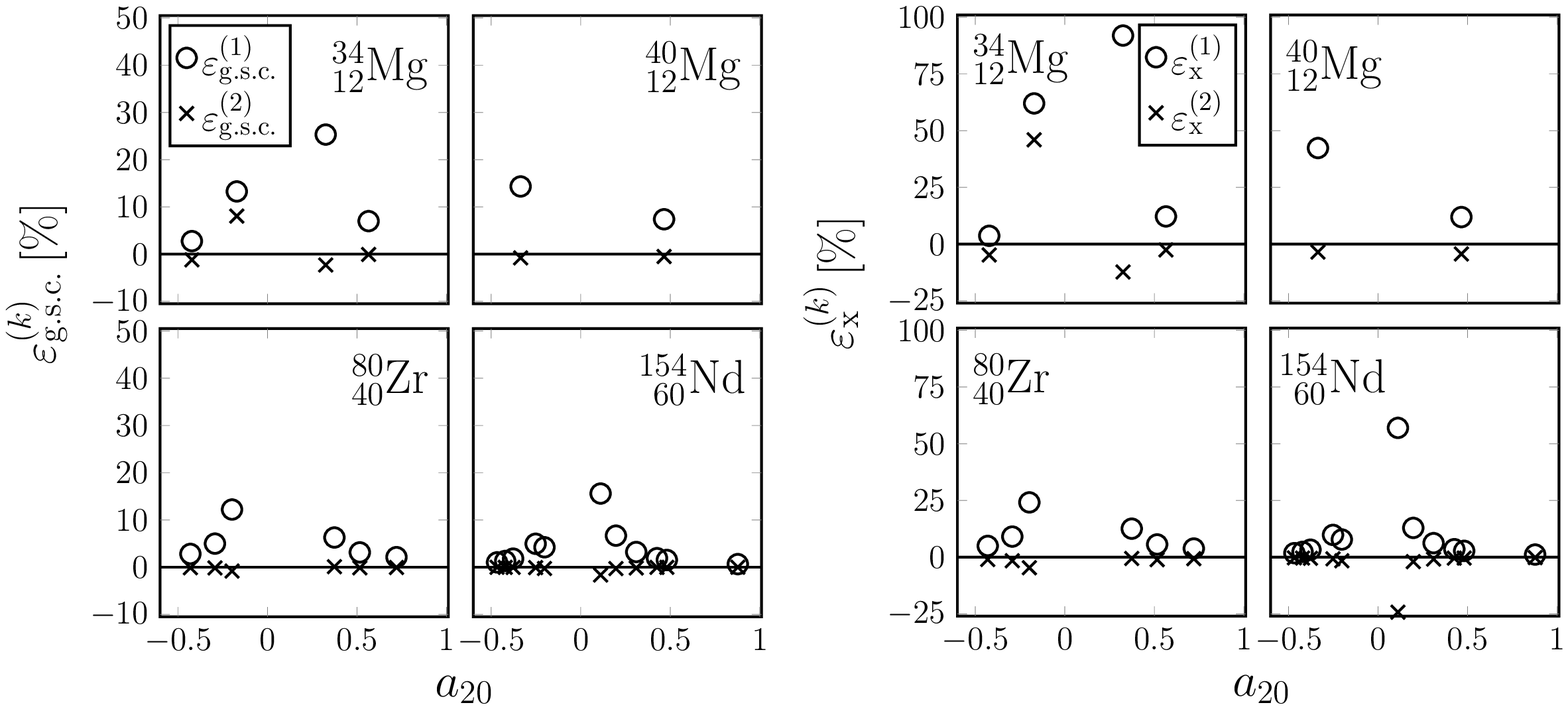}
\caption[]
{
The $a_{20}$ dependence of $\varepsilon^{(k)}_{\mathrm{g.s.c.}}$ and $\varepsilon^{(k)}_{\mathrm{x}}$ for $k=1,2$ in Eq. \eqref{eq:relative_error}.
}
\label{corr_fun_PY}
\end{figure}
\end{widetext}

We next investigate the validity of the approximation in Eq. \eqref{eq:0S0_moi_approx0}.
The values of $s_2$ and $s_4$ are calculated by using Eq. \eqref{eq:Taylor_S01} via numerical differentiation for $\hat{\mathcal{S}}=\hat{H}$.
In Fig. \ref{corr_fun_PY}, the $a_{20}$ dependence of the following quantities is shown (see also Eqs. \eqref{eq:0S0}, \eqref{eq:def_moi}, and \eqref{eq:g.s.c.}),
\begin{subequations}\label{eq:relative_error}
\begin{align}
&\varepsilon^{(k)}_{\mathrm{g.s.c.}}
:=\,
\frac{ \displaystyle -\sum_{n=1}^{k} s_{2n} \varLambda_{2n} - \varDelta E_{\mathrm{g.s.c.}}}
{\varDelta E_{\mathrm{g.s.c.}}},
\label{eq:relative_error_g.s.c.}\\
&\varepsilon^{(k)}_{\mathrm{x}}
:=\,
\frac{ \displaystyle 3 \sum_{n=1}^{k} s_{2n} \left[ -\frac{1}{2} ( \varLambda_{2n+2} - \varLambda_{2n} \varLambda_{2}) \right] - E_{\mathrm{x}}(2^+)}
{E_{\mathrm{x}}(2^+)},
\label{eq:relative_error_Ex}
\end{align}
\end{subequations}
for $k=1,2$.
At well-deformed minima of $^{80}_{40}$Zr and $^{154}_{~60}$Nd nuclei, both $\varepsilon^{(1)}_{\mathrm{g.s.c.}}$ and $\varepsilon^{(1)}_{\mathrm{x}}$ are less than a few percents.
However, they are large for $^{34,40}_{~~~12}$Mg nuclei or weakly-deformed minima.
Regardless of nuclides, $|\varepsilon^{(2)}_{\mathrm{g.s.c.}}|$ is smaller than $|\varepsilon^{(1)}_{\mathrm{g.s.c.}}|$.
Except for $a_{20}=-0.42$ of $^{34}_{12}$Mg, $|\varepsilon^{(2)}_{\mathrm{x}}|$ is also smaller than $|\varepsilon^{(1)}_{\mathrm{x}}|$.
The contributions of the $s_4$ terms to $\varDelta E_{\mathrm{g.s.c.}}$ and $E_{\mathrm{x}}(2^+)$, thereby to the moment-of-inertia, turn out to be significant for $^{34,40}_{~~~12}$Mg nuclei or weakly-deformed minima.

\subsection{Angle dependence of overlap function}
\label{subsec:angle_dep}

In this subsection, the dependence of the overlap functions $\ev*{ e^{-i\hat{J}_y \beta} }{ \Phi_0 }$ and $\mathcal{S}^{01}(\beta)$ in Eq. \eqref{eq:def_S01} on the angle $\beta$ are discussed for further understanding of the numerical results in Sections \ref{subsec:contribution} and \ref{subsec:influence}, such as the $J(J+1)$ rule and the ratio
$\mathcal{S}_{\mathrm{x}}(2^+) / E_{\mathrm{x}}(2^+)$.

\begin{figure}[H]
\centering
\includegraphics[width=8.6cm,clip]{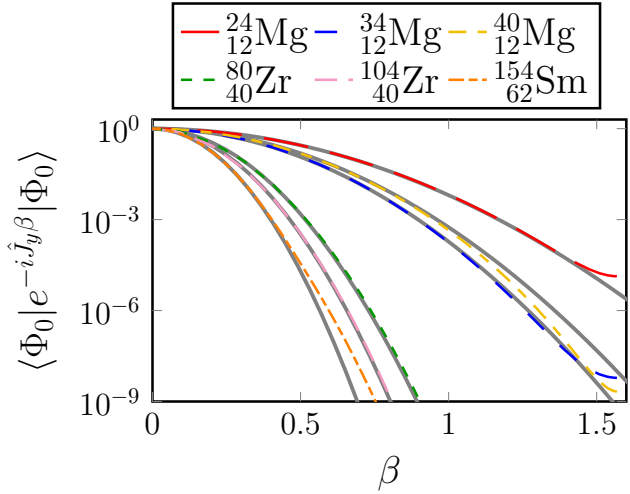}
\caption[]
{
The overlap functions $\ev*{ e^{-i\hat{J}_y \beta} }{ \Phi_0 }$ for the deformed nuclei at their lowest minima.
Gray lines are obtained by the Gaussian approximation in Eq. \eqref{eq:onishi_approx}.
}
\label{onishi_pro}
\end{figure}

In Fig. \ref{onishi_pro}, the overlap functions $\ev*{ e^{-i\hat{J}_y \beta} }{ \Phi_0 }$ are shown for the deformed nuclei at their lowest minima.
The overlap functions for the $^{80}_{40}$Zr and $^{154}_{~62}$Sm nuclei have sharper peaks than those for the $^{34,40}_{~~~12}$Mg nuclei.
The fluctuation $\sigma [ \hat{J}_y ]$ is connected to the coefficient of the second derivative of $\ev*{ e^{-i\hat{J}_y \beta} }{ \Phi_0 }$ at $\beta=0$ via Eq. \eqref{eq:sigma}.
The values of $(\sigma [ \hat{J}_y ])^2$ for $^{24}_{12}$Mg, $^{34}_{12}$Mg, $^{80}_{40}$Zr, $^{104}_{~40}$Zr, and $^{154}_{~62}$Sm, which are calculated from $\ev*{ \hat{\bm{J}}^{\,2} }{ \Phi_0 }$ (see Eq. \eqref{eq:sigma_J_y^2_J^2}), are 10.1, 17.2, 51.9, 64.2, and 86.6, respectively.
The Gaussian approximation in Eq. \eqref{eq:onishi_approx} holds well except at $\beta \approx \pi/2$, although the overlap functions for $_{12}$Mg nuclei have broad peak.
Recall that $(\sigma [ \hat{J}_y ])^{-1}$ is the width of the Gaussian in this approximation.

\begin{figure}[H]
\centering
\includegraphics[width=8.6cm,clip]{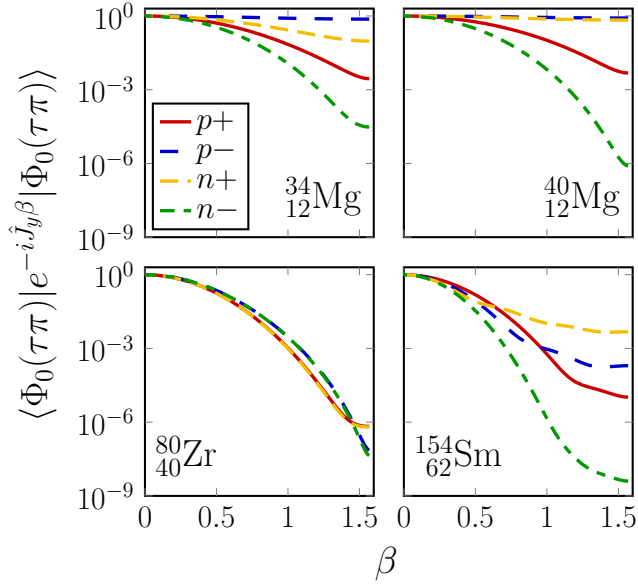}
\caption[]
{
$\ev*{ e^{-i\hat{J}_y \beta} }{ \Phi_0 (\tau \pi) } (\tau=p,n, \pi=+,-)$ for the deformed $^{34}_{12}$Mg, $^{40}_{12}$Mg, $^{80}_{40}$Zr, and $^{154}_{~62}$Sm nuclei at their lowest minima.
}
\label{onishi_parity}
\end{figure}

In the present AMP calculations, the overlap function $\ev*{ e^{-i\hat{J}_y \beta} }{ \Phi_0 }$ can be factorized via isospin and parity as Eq. \eqref{eq:phi_tz_pi} because the rotational operator $e^{-i \hat{J}_y \beta}$ does not mix isospin and parity.
Figure \ref{onishi_parity} shows the components of the overlap functions $\ev*{ e^{-i\hat{J}_y \beta} }{ \Phi_0 (\tau \pi) }$ ($\tau=p,n$ and $\pi=+,-$), for the deformed $^{34}_{12}$Mg, $^{40}_{12}$Mg, $^{80}_{40}$Zr, and $^{154}_{~62}$Sm nuclei at their lowest minima.
For the $_{12}$Mg nuclei, some components have almost spherical structure and hardly depend on the $\beta$ angle.
For the $^{40}_{12}$Mg nucleus, the $n-$ component is well-deformed, which may be related to the deformed halo with peanut shape \cite{Na08}.
For the $^{80}_{40}$Zr nucleus, all $\tau \pi$ components are similarly deformed.
For the $^{154}_{~62}$Sm nucleus, the $n-$ component is strongly deformed, though the others are not so strongly deformed.

\begin{widetext}
\begin{figure}[H]
\centering
\includegraphics[width=17.8cm,clip]{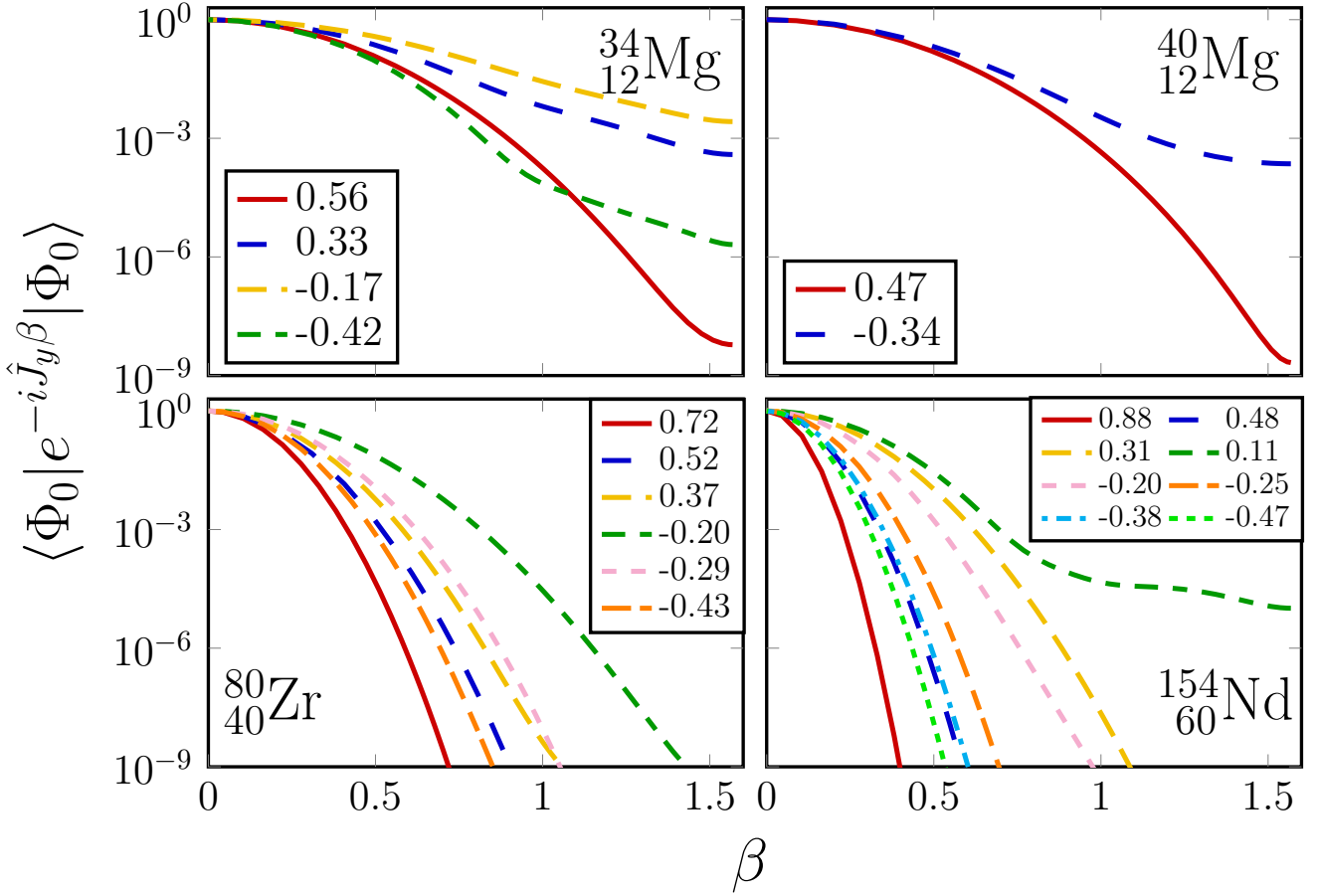}
\caption[]
{
The overlap functions $\ev*{ e^{-i\hat{J}_y \beta} }{ \Phi_0 }$ for the $^{34}_{12}$Mg, $^{40}_{12}$Mg, $^{80}_{40}$Zr, and $^{154}_{~60}$Nd nuclei, including their local minima.
The individual lines correspond to the various $a_{20}$ values shown in the insets.
}
\label{onishi_def_para}
\end{figure}
\end{widetext}

Figure \ref{onishi_def_para} shows $\ev*{ e^{-i\hat{J}_y \beta} }{ \Phi_0 }$ for minima of the $^{34}_{12}$Mg, $^{40}_{12}$Mg, $^{80}_{40}$Zr, and $^{154}_{~60}$Nd nuclei, including their local minima with various $a_{20}$ values.
The overlap functions have sharper peaks near $\beta =0$ irrespective of nuclides as $|a_{20}|$ increases.
A similar result is obtained in Ref. \cite{BB21}.
The sharpness of the peak near $\beta =0$ of $\ev*{ e^{-i\hat{J}_y \beta} }{ \Phi_0 }$ corresponds with the fluctuation $\sigma[\hat{J}_y]$.
The large fluctuation takes place when $|a_{20}|$ is large for the heavy nuclei \cite{RS80}, which leads to a sharp peak near $\beta=0$.
The overlap function depends on the mass number as well as on $a_{20}$.
The Gaussian approximation sometimes fails for the $_{12}$Mg nuclei or the weakly-deformed minima.
As $\varLambda_{2n}$ in Eq. \eqref{eq:N_2n_Lambda_2n} is determined by only $\ev*{ e^{-i\hat{J}_y \beta} }{ \Phi_0 }$, it is fair to say that the results in Fig. \ref{Lambda_2n} originates from those in Fig. \ref{onishi_def_para}.

\begin{figure}[H]
\centering
\includegraphics[width=8.6cm,clip]{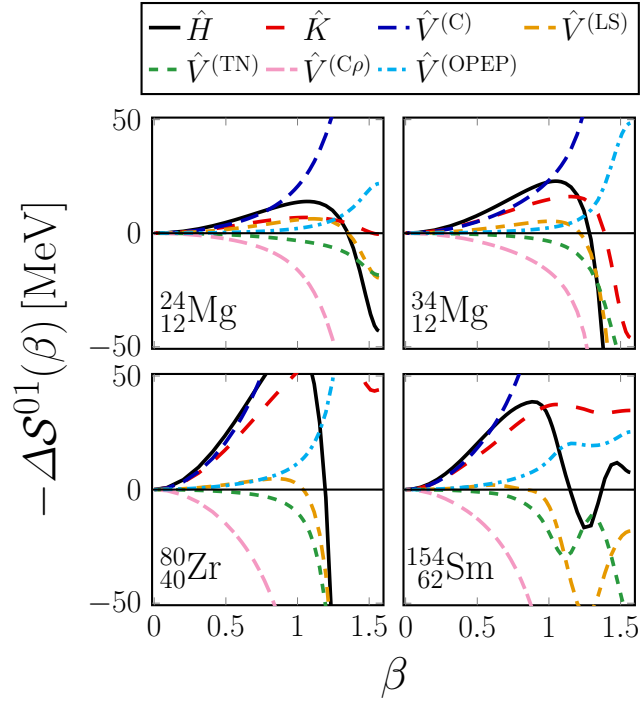}
\caption[]
{
The behavior of $-\varDelta \mathcal{S}^{01}(\beta)$ for the individual terms of the Hamiltonian for the deformed $^{24}_{12}$Mg, $^{34}_{12}$Mg, $^{80}_{40}$Zr, and $^{154}_{~62}$Sm nuclei at their lowest minima.
}
\label{beta_dep_int_all}
\end{figure}

We show $-\varDelta \mathcal{S}^{01}(\beta)$ which is defined by
\begin{equation}\label{eq:Delta_S01}
\varDelta \mathcal{S}^{01}(\beta)
:=
\mathcal{S}^{01}(\beta)-\mathcal{S}^{01}(\beta=0),
\end{equation}
for the deformed $^{24}_{12}$Mg, $^{34}_{12}$Mg, $^{80}_{40}$Zr, and $^{154}_{~62}$Sm nuclei at their lowest minima in Fig. \ref{beta_dep_int_all}.
These results are related to those in Fig. \ref{percentage_HF_pro}.
As in Eq. \eqref{eq:C}, we have
\begin{equation}\label{eq:DeltaS01_correlation}
\left. -\, \frac{d^{\,2}}{d\beta^{\,2}}\,
\varDelta \mathcal{S}^{01}(\beta) \right| _{\beta=0}
=
C[ \hat{\mathcal{S}}, \hat{J}_y^{\,2} ].
\end{equation}
The values of $C[ \hat{\mathcal{S}}, \hat{J}_y^{\,2} ]$ for $\hat{\mathcal{S}}=\hat{K}$, $\hat{V}^{\mathrm{(C)}}$, and $\hat{V}^{\mathrm{(LS)}}$ are positive, and those for $\hat{V}^{\mathrm{(TN)}}$ and $\hat{V}^{\mathrm{(C\rho)}}$ are negative.
As the mass number increases, the values of $|C[ \hat{\mathcal{S}}, \hat{J}_y^{\,2} ]|$ significantly increase except for $\hat{\mathcal{S}}=\hat{V}^{\mathrm{(LS)}}$, $\hat{V}^{\mathrm{(TN)}}$, and $\hat{V}^{\mathrm{(OPEP)}}$.
Although they are not shown, $-\varDelta \mathcal{S}^{01}(\beta) \approx 0$ and $C[\hat{\mathcal{S}},\hat{J}_y^{\,2}]\approx0$ for $\hat{V}_{\mathrm{Coulomb}}$ and $\hat{H}_{\mathrm{c.m.}}$, independent of nuclides.
The values of $-\varDelta \mathcal{S}^{01}(\beta)$ far from $\beta \approx 0$ strongly depend on nuclides, which are influenced by the higher-order terms of the cumulant expansion in Eq. \eqref{eq:S01_s2n}.

\begin{figure}[H]
\centering
\includegraphics[width=8.6cm,clip]{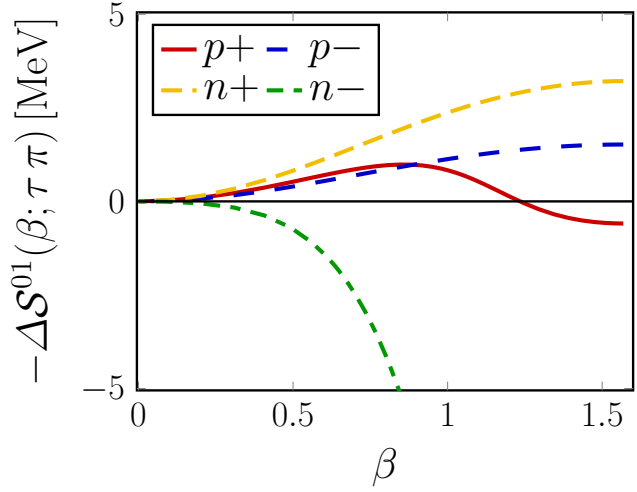}
\caption[]
{
The function $-\varDelta \mathcal{S}^{01}(\beta;\tau \, \pi)$ in Eq. \eqref{eq:S01_taupi_all} for $\hat{\mathcal{S}}=\hat{K}$ at the oblate minimum of $^{40}_{12}$Mg.
}
\label{beta_dep_kinetic_parity}
\end{figure}

In the present work, the values of $C[\hat{\mathcal{S}}, \hat{J}_y^{\,2}]$ for $\hat{\mathcal{S}}=\hat{K}$ are almost always positive.
However, there exists an exception; the local oblate minimum of $^{40}_{12}$Mg.
We decompose $-\varDelta \mathcal{S}^{01}(\beta)$ for $\hat{K}$ as
\begin{subequations}\label{eq:S01_taupi_all}
\begin{align}
\mathcal{S}^{01}(\beta;\tau \, \pi)
:=&\,
\frac{ \ev*{ \hat{\mathcal{S}} \, e^{-i \hat{J}_y \beta} }{ \Phi_0 (\tau \pi) } }
{ \ev*{ e^{-i \hat{J}_y \beta} }{ \Phi_0 } },
\label{eq:S01_taupi}\\
\varDelta \mathcal{S}^{01}(\beta;\tau \, \pi)
:=&\,
\mathcal{S}^{01}(\beta;\tau \, \pi)-\mathcal{S}^{01}(\beta=0;\tau \, \pi),
\label{eq:Delta_S01_taupi}
\end{align}
\end{subequations}
for $\tau=p,n$ and $\pi=+,-$, and show $-\varDelta \mathcal{S}^{01}(\beta;\tau \, \pi)$ in Fig. \ref{beta_dep_kinetic_parity} for the oblate minimum of $^{40}_{12}$Mg.
The curvature of $-\varDelta \mathcal{S}^{01}(\beta;n \, -)$ at $\beta \approx 0$ is negative and significant although those for the others are positive and small.
This anomalous result is related to the negative contribution of $\hat{K}$ exhibited in Fig. \ref{percentage_def_para}.

\subsection{Comparison of $E_{\mathrm{x}}(2^+)$ with rigid-rotor model and experiment}

\begin{widetext}
\begin{figure}[H]
\centering
\includegraphics[width=17.8cm,clip]{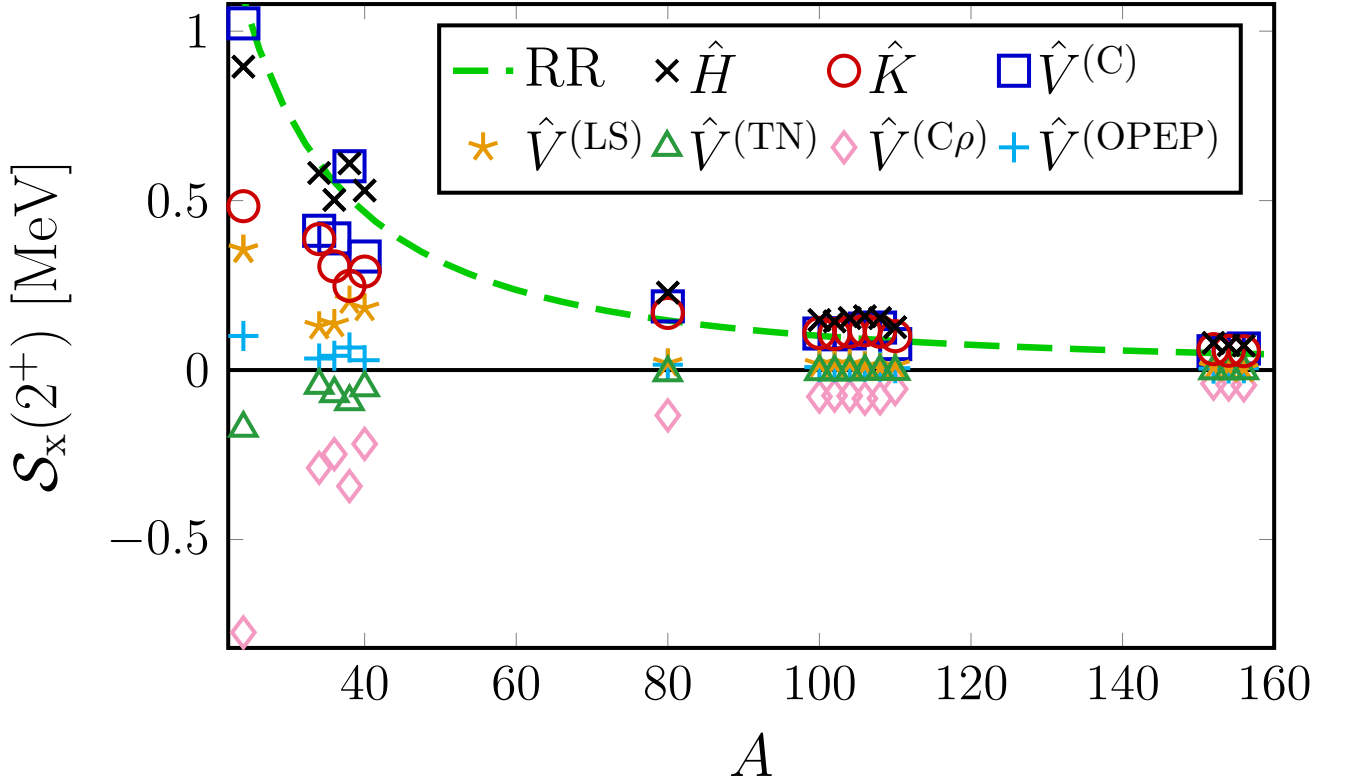}
\caption[]
{
The excitation energies $E_{\mathrm{x}}(2^+)$ for the deformed $^{24,34-40}_{~~~~~~~\,12}$Mg, $^{80,100-110}_{~~~~~~~~~\,40}$Zr, and $^{152-156}_{~~~~~~\,62}$Sm isotopes at their lowest minima.
The black cross symbols represent the values of $E_{\mathrm{x}}(2^+)$ in the present calculations.
The corresponding values of $\mathcal{S}_{\mathrm{x}}(2^+)$ are also shown.
The green dashed line is the rigid-rotor value in Eq. \eqref{eq:RR_approx} \cite{BM98}.
}
\label{exc_A_HF}
\end{figure}
\end{widetext}

Figure \ref{exc_A_HF} shows the calculated excitation energies $E_{\mathrm{x}}(2^+)$ and $\mathcal{S}_{\mathrm{x}}(2^+)$, the latter of which is the contribution of the individual terms of the effective Hamiltonian to the rotational energies (see Eq. \eqref{eq:S_x}), for the deformed $_{12}$Mg, $_{40}$Zr, and $_{62}$Sm isotopes at their lowest minima.
As expected, the calculated $E_{\mathrm{x}}(2^+)$ tends to decrease as $A$ increases.
The absolute values of $\mathcal{S}_{\mathrm{x}}(2^+)$ for the individual terms of the effective Hamiltonian do, as well.
The rigid-rotor value \cite{BM98} is also shown,
\begin{equation}\label{eq:RR_approx}
\begin{split}
&E_{\mathrm{x}}^{(\mathrm{RR})}(J^+)
=\,
\frac{J(J+1)}{2 \, \mathcal{I}^{\mathrm{(RR)}}},\\
&\mathcal{I}^{\mathrm{(RR)}}
\approx\,
0.0138 \, A^{5/3} [\mathrm{MeV}^{-1}].
\end{split}
\end{equation}
In the classical mechanics, the rotational energy of the rigid body comes from kinetic energy.
Interestingly, the values of $\mathcal{S}_{\mathrm{x}}(2^+)$ for $\hat{\mathcal{S}}=\hat{K}$ are close to the rigid-rotor value in the $_{40}$Zr and $_{62}$Sm regions.

\begin{figure}[H]
\centering
\includegraphics[width=8.6cm,clip]{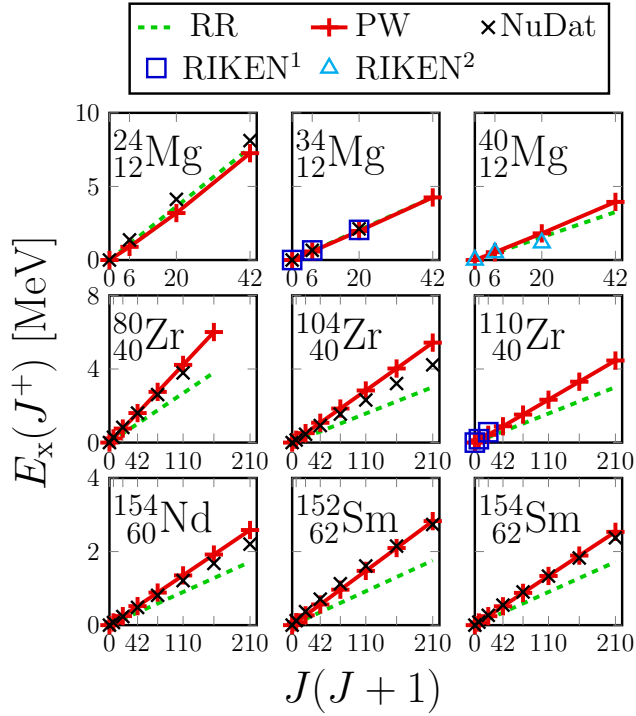}
\caption[]
{
The excitation energies for the deformed nuclei at their lowest minima.
The horizontal axis is the value of $J(J+1)$ with the non-negative integer $J$.
The experimental values displayed by black cross \cite{NNDC}, blue square \cite{DS13}, and sky-blue triangle symbols \cite{CF19}.
The green dashed lines are the rigid-rotor values calculated from Eq. \eqref{eq:RR_approx}.
The red plus symbols are obtained by the present work.
}
\label{energy_group}
\end{figure}

Figure \ref{energy_group} shows the excitation energies for the deformed nuclei at their lowest minima, all of which have prolate shapes.
The rigid-rotor energies are low compared to the experimental ones, {\it e.g.}, for the
$_{40}$Zr, $_{60}$Nd, and $_{62}$Sm regions.
The excitation energies obtained by the present calculations are close to the experimental values of $E_{\mathrm{x}}(2^+)$ for all nuclides.
However, we should be careful in comparing the values obtained by the present AMP calculations with those of the experiment.
The pair correlations will reduce the moment-of-inertia and raise the excitation energies \cite{Be59}, while the intrinsic state is not always stable for increasing $J$, tending to decrease the excitation energies.
It should also be noted that there is uncertainty in treating the density-dependent terms in the AMP calculations.

\section{Conclusion}

The Peierls-Yoccoz (PY) rotational energy of nuclei has been analyzed by the AMP calculation for the self-consistent axial-HF solutions, using the semi-realistic effective Hamiltonian M3Y-P6.
The contributions of the individual terms of the Hamiltonian to the rotational energies have been analyzed.
Except for the light nuclei or the weakly-deformed solutions, their ratios are insensitive to nuclides and states.
The contributions of the kinetic energies are large and close to the rigid-rotor values.
A large cancellation occurs between the density-dependent channel and the density-independent one in the central force, although their sum is still sizable.
The contributions of the noncentral forces are small.
In contrast, the results significantly depend on nuclei and deformation for the light nuclei or the weakly-deformed solutions.
The contributions of the noncentral forces are not negligible.
Regardless of nuclides, the attractive forces decrease the moment-of-inertia, and the repulsive forces increase it.
The pair correlations and the $J$-dependence of the intrinsic state may influence the results for actual nuclei, and we leave them for future works.

By using the cumulant expansion, a general formula for the PY rotational energy is derived on the basis of the AMP.
This formula is a generalization of those in Refs. \cite{PY57,Yo57,Ve63,Ve64,Ka68,RS80}.
It is suggested that the newly found higher-order terms of the cumulant expansion play roles in the light nuclei or the weakly-deformed solutions, contributing to the rotational energy.

\section*{acknowledgments}

The authors are grateful to H. Kurasawa and S. Iwasaki for discussions.
In this research, the numerical calculations were carried out on Yukawa-21 at YITP in Kyoto University.
This research also used computational resources of Oakforest PACS provided by the Multidisciplinary Cooperative Research Program in Center for Computational Sciences, University of Tsukuba, and HITACHI SR24000 at the Institute of Management and Information Technologies, Chiba University.
This research had been supported by the research assistant program at Chiba University.
We thank K. Neergard for drawing our attention to references proving the non-negativity of matrices in Appendix \ref{ap:nonneg}.

\appendix
\section{Gaussian approximation connected to AMP}
\label{ap:Gauss_approx}

In this appendix, the Gaussian approximation \cite{Yo57,Ve64,Ka68,RS80} for the rotational energy in Sec. \ref{subsec:AMP_rot} is discussed with higher-$s_{2n}$-terms in Eq. \eqref{eq:SJN_expand2}.
There are certain cases that the overlap function $\ev*{ e^{-i \hat{J}_y \beta} }{ \Phi_0 }$ is well approximated by the Gaussian function as
\begin{equation}\label{eq:onishi_approx}
\ev*{ e^{-i \hat{J}_y \beta} }{ \Phi_0 }
=\,
1
-\frac{1}{2} \ev*{ \hat{J}_y^{\,2} }{ \Phi_0 } \beta^2
+
\cdots
\approx\,
e^{-\frac{1}{2} (\sigma [ \hat{J}_y ])^2 \beta^2 }.\\
\end{equation}
The width $(\sigma [ \hat{J}_y ])^{-1}$ is not always narrow.
For $x>0$, the following functions are defined,
\begin{subequations}\label{eq:N_2n^G_Lambda_2n^G}
\begin{align}
N_{2n}^{\mathrm{(G)}}(x)
:=&\,
\displaystyle \int_0^{\pi/2} d\beta
\sin \beta
\, \beta^{2n}
e^{-\frac{1}{2} x \beta^2 },
\label{eq:N_2n^G}\\
\varLambda_{2n}^{\mathrm{(G)}}(x)
:=&\,
\frac{N_{2n}^{\mathrm{(G)}}(x)}{N_0^{\mathrm{(G)}}(x)},
\label{eq:Lambda_2n^G}
\end{align}
\end{subequations}
analogously to Eq. \eqref{eq:N_2n_Lambda_2n}.
The function $\varLambda_{2}^{\mathrm{(G)}}(x)$ is called universal function in Ref. \cite{LRV04}.
\begin{figure}[H]
\centering
\includegraphics[width=8.6cm,clip]{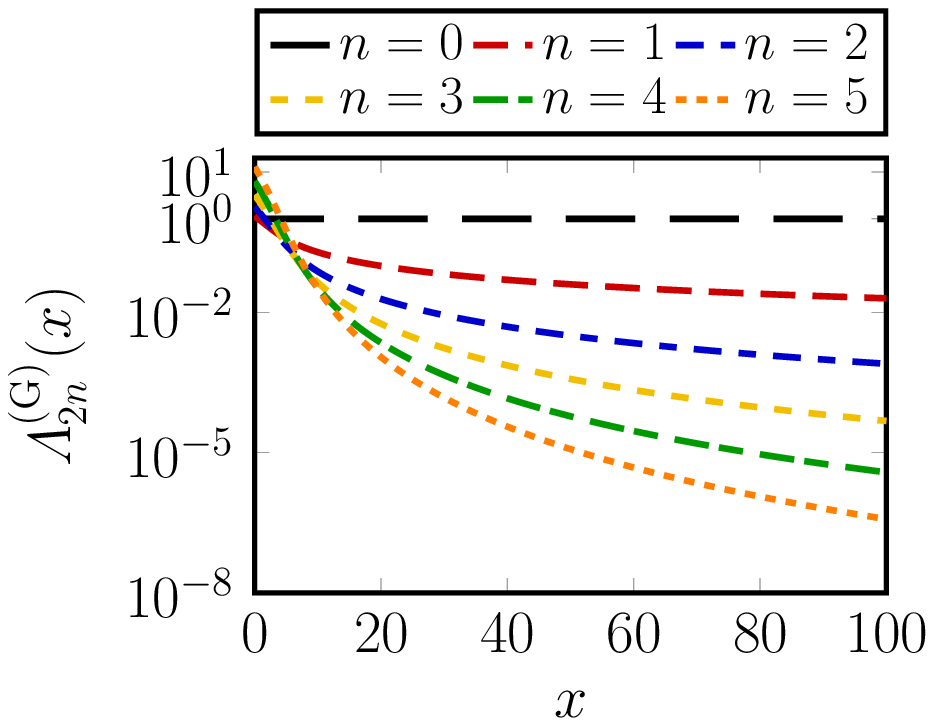}
\caption[]
{
$\varLambda_{2n}^{\mathrm{(G)}}(x)$ for $n=0,...,5$.
}
\label{fig:uf}
\end{figure}
In Fig. \ref{fig:uf}, $\varLambda_{2n}^{\mathrm{(G)}}(x)$ in Eq. \eqref{eq:Lambda_2n^G} is shown.
For small $n$ and large $x$, the following relation is satisfied,
\begin{equation}\label{eq:Lambda_2n_condition1}
\varLambda_{2n}^{\mathrm{(G)}}(x)
>
\varLambda_{2n+2}^{\mathrm{(G)}}(x).
\end{equation}
The recurrence relations of $N_{2n}^{\mathrm{(G)}}(x)$ and $\varLambda_{2n}^{\mathrm{(G)}}(x)$ are as follows,
\begin{subequations}
\begin{align}
\frac{d}{dx} \, N_{2n}^{\mathrm{(G)}}(x)
=&
-\frac{1}{2} \, N_{2n+2}^{\mathrm{(G)}}(x),
\label{eq:recurrence_N}\\
\frac{d}{dx} \, \varLambda_{2n}^{\mathrm{(G)}}(x)
=&
-\frac{1}{2} 
\left[ 
\varLambda_{2n+2}^{\mathrm{(G)}}(x)
-
\varLambda_{2n}^{\mathrm{(G)}}(x) 
\varLambda_{2}^{\mathrm{(G)}}(x) 
\right].
\label{eq:recurrence_Lambda}
\end{align}
\end{subequations}
Equations \eqref{eq:0S0} and \eqref{eq:def_moi} are approximated as follows,
\begin{subequations}\label{eq:0S0_moi_approx1}
\begin{align}
&\ev*{ \hat{\mathcal{S}} }{ 0 }
\approx
\left. \sum_{n=0}^{\infty} s_{2n} \varLambda_{2n}^{\mathrm{(G)}}(x)
\right|_{x=(\sigma [ \hat{J}_y ])^2},
\label{eq:0S0_approx1} \\
&\frac{1}{\mathcal{I}[\hat{\mathcal{S}}]}
\approx
\left. \sum_{n=1}^{\infty} s_{2n} \, \frac{d}{dx} \, \varLambda_{2n}^{\mathrm{(G)}}(x) \right|_{x=(\sigma [ \hat{J}_y ])^2},
\label{eq:moi_approx1}
\end{align}
\end{subequations}
regardless of the value of $(\sigma [ \hat{J}_y ])^2$.

If the width of the Gaussian $(\sigma [ \hat{J}_y ])^{-1}$ is narrow enough, $N_{2n}^{\mathrm{(G)}}(x)$ is approximated by taking $\sin\beta \approx \beta$ in Eq. \eqref{eq:N_2n^G},
\begin{equation}\label{eq:N2n_sin}
N_{2n}^{\mathrm{(G)}}(x)
\approx
\int_{0}^{\pi/2} d\beta \, \beta^{2n+1} e^{-\frac{1}{2} x \beta^2 }
=
\, 2^n \int_0^{\lambda} du \, u^n e^{-xu},
\end{equation}
where $\lambda := \pi^2 /8$.
The recurrence relation in Eq. \eqref{eq:recurrence_N} is satisfied for the approximate $N_{2n}^{\mathrm{(G)}}(x)$ in Eq. \eqref{eq:N2n_sin}.
For $n=0$, the integration in Eq. \eqref{eq:N2n_sin} can be done analytically,
\begin{equation}\label{eq:N0_sin}
N_{0}^{\mathrm{(G)}}(x)
\approx
\frac{1}{x} \, (1-e^{-\lambda x}).
\end{equation}
By using Eqs. \eqref{eq:recurrence_N} and \eqref{eq:N0_sin}, an analytical expression of $N_{2n}^{\mathrm{(G)}}(x)$ is obtained,
\begin{equation}\label{eq:N2n_sin_analytical}
N_{2n}^{\mathrm{(G)}}(x)
\approx
\frac{2^n n!}{x^{n+1}}
\left(
1-e^{-\lambda x}
\displaystyle{
\sum_{m=0}^{n} \frac{(\lambda x)^m}{m!}
}
\right).
\end{equation}

By widening the range of integral $\pi/2 \rightarrow \infty$ ({\it i.e.}, $\lambda \rightarrow \infty$) in Eq. \eqref{eq:N2n_sin}, the following equation is obtained,
\begin{equation}\label{eq:N2n_sin_inf}
N_{2n}^{\mathrm{(G)}}(x)
\approx \,
\frac{2^n n!}{x^{n+1}},
\end{equation}
and,
\begin{equation}\label{eq:lambda2n_sin_inf}
\varLambda_{2n}^{\mathrm{(G)}}(x)
\approx
\frac{2^n n!}{x^{n}}.
\end{equation}
Equation \eqref{eq:lambda2n_sin_inf} satisfies Eq. \eqref{eq:Lambda_2n_condition1} for small $n$ and large $x$.
Although Eq. \eqref{eq:Lambda_2n_condition1} breaks down at extremely large $n$ for any $x$, $c_{2n}$ in Eq. \eqref{eq:c2n} eases a problem of convergence in Eq. \eqref{eq:JSJ_Lambda} via,
\begin{equation}
\ev*{ (\hat{J}_{+}-\hat{J}_{-})^{2n} }{ J0 }
\sim
\begin{pmatrix}
2n\\
n
\end{pmatrix}
(-)^n J^{2n},
\end{equation}
and,
\begin{equation}
c_{2n} \, \varLambda^\mathrm{(G)}_{2n}
\sim
\frac{1}{n!}
\left(
-\frac{J^{\, 2}}{2x}
\right)^n.
\end{equation}
If we neglect higher-$s_{2n}$-terms in Eq. \eqref{eq:0S0_moi_approx1},
$\ev*{ \hat{\mathcal{S}} }{ 0 }$ and $\mathcal{I}[\hat{\mathcal{S}}]$ are approximated by using Eq. \eqref{eq:lambda2n_sin_inf},
\begin{subequations}\label{eq:0S0_moi_approx2}
\begin{align}
\ev*{ \hat{\mathcal{S}} }{ 0 }
\approx&
\ev*{ \hat{\mathcal{S}} }{ \Phi_0 }
-\frac{ C[ \hat{\mathcal{S}}, \hat{J}_y^{\,2} ] }
{ (\sigma [ \hat{J}_y ])^2 },
\label{eq:0S0_approx2}\\
\frac{1}{\mathcal{I}[\hat{\mathcal{S}}]}
\approx& \,
\frac{C[ \hat{\mathcal{S}}, \hat{J}_y^{\,2} ]}
{(\sigma[\hat{J}_y])^{4}}.
\label{eq:moi_approx2}
\end{align}
\end{subequations}
Equation \eqref{eq:0S0_moi_approx2} is the result of the Kamlah expansion \cite{Ka68,RS80}, and Eq. \eqref{eq:moi_approx2} is the Yoccoz moment-of-inertia \cite{Yo57,Ve64,RS80}.

\section{AMP for non-orthogonal bases}\label{ap:AMP}

A summary of the MF theory for non-orthogonal bases, particularly the HFB theory, is given in the appendix of Ref. \cite{Na06}.
In this appendix, we present a part of the method of AMP that is characteristic of the non-orthogonal bases.
The single-particle (s.p.) base ket is represented by $\ket{k}$, and $\mathsf{N}_{kk'}:=\braket*{k}{k'}$ is the norm matrix.
We assume that $\mathsf{N}$ is positive definite, then the completeness holds, $\sum_{kk'} \ket{k} \left( \mathsf{N}^{-1} \right)_{k k'} \bra{k'} = \hat{1}$, where $\hat{1}$ is the identity operator in the s.p. space.
We denote the creation (annihilation) operator for the s.p. basis $k$ by $c^{\dagger}_k$ ($c_k$).
They obey the fermionic anti-commutation relations,
\begin{equation}\label{eq:antic}
\{ c_k, c^{\dagger}_{k'} \} = \mathsf{N}_{kk'} ,~~~~~
\{ c_k, c_{k'} \} =0,~~~~~
\{ c^{\dagger}_k, c^{\dagger}_{k'} \} =0.
\end{equation}
The particle vacuum $\ket*{0}_c$ is defined by $c _k \ket*{0}_c = 0$ for all $k$, which satisfies ${}_{c}\!\braket{ 0 }{ 0 }\!{}_{c} =1$.

The generalized Bogolyubov transformation is given as \cite{RS80}
\begin{equation}\label{eq:transUV}
\begin{split}
\alpha^{\dagger}_i
:=
\sum_{k=1}^{M}
\left( c^{\dagger}_k \mathsf{U}_{ki} + c_k \mathsf{V}_{ki} \right), \\
\alpha_i
:=
\sum_{k=1}^{M}
\left( c_k \mathsf{U}^{\ast}_{ki} + c^{\dagger}_k \mathsf{V}^{\ast}_{ki} \right),
\end{split}
\end{equation}
where $M$ is the number of the bases, the matrices $\mathsf{U}$ and $\mathsf{V}$ are $M \times M$ square matrices.
In the vector and matrix representation, Eq. \eqref{eq:transUV} can be expressed as follows,
\begin{equation}\label{eq:transW}
( \boldsymbol \alpha^{\dagger} \,\,\,\,\, \boldsymbol{\alpha} )
=
( \bold{c}^{\dagger} \,\,\,\,\, \bold{c} )
\mathsf{W}
,~~~~~~
\mathsf{W}
:=
\begin{pmatrix}
\mathsf{U} & \mathsf{V}^{\ast} \\
\mathsf{V} & \mathsf{U}^{\ast}
\end{pmatrix},
\end{equation}
where $( \bold{c}^{\dagger} \,\,\,\,\, \bold{c} )$ represents $( c_1^{\dagger} \cdots c_M^{\dagger} \, c_1 \cdots c_M )$.
In contrast to $c^{\dagger}_k$ and $c_k$, $\alpha^{\dagger}_i$ and $\alpha_i$ obey the usual fermionic canonical anti-commutation relations,
\begin{equation}\label{eq:antialpha}
\{ \alpha_i, \alpha^{\dagger}_{i'} \} = \delta_{ii'} ,~~~~~
\{ \alpha_i, \alpha_{i'} \} =0,~~~~~
\{ \alpha^{\dagger}_i, \alpha^{\dagger}_{i'} \} =0.
\end{equation}
The matrix $\mathsf{W}$ satisfies the following equation,
\begin{equation}\label{eq:WNW}
\mathsf{W}^{\dagger} \mathsf{N}' \mathsf{W}=1
,\quad
\mathsf{N}':=
\begin{pmatrix}
\mathsf{N} & \mathsf{0} \\
\mathsf{0} & \mathsf{N}^{\ast}
\end{pmatrix}.
\end{equation}
The HFB vacuum $\ket{\Phi}$ is defined by $\alpha_i \ket{\Phi} = 0$ for all $i$, and satisfies $\braket{ \Phi }{ \Phi } =1$.

The transformation by the rotational operator $e^{-i\hat{J}_y\beta}$ for the s.p. bases is represented as follows,
\begin{equation}\label{eq:def_D}
e^{-i\hat{J}_y\beta}
( \bold{c}^{\dagger} \,\,\,\,\, \bold{c} )
e^{i\hat{J}_y\beta}
=:
( \bold{c}^{\dagger} \,\,\,\,\, \bold{c} ) \mathsf{D}'
,\quad
\mathsf{D}'
=
\begin{pmatrix}
\mathsf{D} & \mathsf{0} \\
\mathsf{0} & \mathsf{D}^{\ast}
\end{pmatrix}.
\end{equation}
In the case of the spherically symmetric s.p. bases $k=( \nu \ell j m t_z)$, as in the GEM of Refs. \cite{NS02,Na08},
\begin{equation}\label{eq:DcD}
e^{-i\hat{J}_y\beta}
c^{\dagger}_{ \nu \ell j m t_z}
e^{i\hat{J}_y\beta}
=
\sum_{ m' }
c^{\dagger}_{ \nu \ell j m' t_z}
d^{(j)}_{ m' m }(\beta),
\end{equation}
and the matrix elements of $\mathsf{D}$ in Eq. \eqref{eq:def_D} are $\mathsf{D}_{kk'} = \delta_{\nu \nu'} \delta_{\ell \ell'} \delta_{j j'} \delta_{t_z t_z'} d^{(j)}_{ m m' }(\beta)$ .
We define a matrix $\mathsf{T}$,
\begin{equation}\label{eq:T_mat}
e^{-i\hat{J}_y\beta}
(\boldsymbol \alpha^{\dagger} \,\,\,\,\, \boldsymbol \alpha)
e^{i\hat{J}_y\beta}
=:
(\boldsymbol \alpha^{\dagger} \,\,\,\,\, \boldsymbol \alpha) \mathsf{T}
,\quad
\mathsf{T}
=
\begin{pmatrix}
\mathsf{T_{11}} & \mathsf{T_{12}}\\
\mathsf{T_{21}} & \mathsf{T_{22}}\\
\end{pmatrix},
\end{equation}
which satisfies the following relation,
\begin{equation}\label{eq:WNDW}
\mathsf{T}
=
\mathsf{W}^{-1} \mathsf{D' W = W^{\dagger} N' D' W},
\end{equation}
with $\mathsf{T_{22}}=\mathsf{T_{11}}^*$ and $\mathsf{T_{21}}=\mathsf{T_{12}}^*$.

For simplicity, we express $\ket{0} := \ket{\Phi}$ and $\ket{1} := e^{-i\hat{J}_y \beta} \ket{\Phi}$, and assume that $\braket{0}{1}$ does not vanish.
For the Hamiltonian $\hat{H}$ consists of the 1-body term $\hat{K}$ and the 2-body term $\hat{V}$, we get the following equation by using the generalized Wick's theorem \cite{BB69, RS80},
\begin{widetext}
\begin{equation}\label{eq:H01}
\begin{split}
\frac{ \mel*{ 0 } { \hat{H} } {1 } }
{ \braket*{ 0 }{1 } }
=&
\sum_{k_1 k_2}
\mel*{ k_2 } { \hat{K} } { k_1 }
\rho^{0 1}_{k_1 k_2}
+
\frac{1}{4}
\sum_{k_1 k_2 k_3 k_4}
\mel*{ k_3 k_4 } { \hat{V} } { k_1 k_2 }_{a}
\left(2	\rho^{0 1}_{k_1 k_3}
\rho^{0 1}_{k_2 k_4}
+
\kappa^{0 1}_{k_1 k_2}
\kappa^{1 0 \ast}_{k_3 k_4}
\right),
\end{split}
\end{equation}
\end{widetext}
where the matrix elements of $\hat{V}$ are anti-symmetrized, and we have defined ``{\it generalized density matrix}'' $\rho ^{01}$, and ``{\it generalized pairing tensors}'' $\kappa ^{01}$ and $\kappa ^{10}$,
\begin{subequations}\label{eq:defEDM01}
\begin{align}
( {\mathsf N} \rho ^{01} {\mathsf N} )_{kk'}
&:=\frac{ \mel*{ 0 } { c^{\dagger}_{k'}	c_{k} } { 1 } }
{ \braket*{ 0 }{ 1 } },
\label{eq:defGDM01}\\
( {\mathsf N} \kappa ^{01} {\mathsf N^{\ast}} )_{kk'}
&:=\frac{ \mel*{ 0 } { c_{k'}	c_{k} } { 1 } }
{ \braket*{ 0 }{ 1 } }, \\
( {\mathsf N} \kappa ^{10} {\mathsf N^{\ast}} )_{kk'}
&:=\frac{ \mel{ 1 } { c_{k'}	c_{k} } { 0 } }
{ \braket*{ 1 }{ 0 } }.
\end{align}
\end{subequations}
The matrices $\rho^{01}$, $\kappa ^{01}$, and $ \kappa ^{10}$ in Eq. \eqref{eq:defEDM01} can be expressed as follows \cite{BB69,RS80},
\begin{equation}\label{eq:calEDM01}
\rho^{01}
=
\widetilde{\mathsf{V}}^{\ast} \mathsf{V}^{\mathrm{T}}, ~~~
\kappa^{01}
=
\widetilde{\mathsf{V}}^{\ast} \mathsf{U}^{\mathrm{T}}, ~~~
\kappa^{10}
=
\mathsf{V}^{\ast} \widetilde{\mathsf{U}}^{\mathrm{T}}, ~~~
\end{equation}
with,
\begin{equation}\label{eq:tildeVU}
\widetilde{\mathsf{V}}
:=
\mathsf{V} + \mathsf{U}^{\ast} \mathsf{X}^{\ast}, ~~~
\widetilde{\mathsf{U}}
:=
\mathsf{U} + \mathsf{V}^{\ast} \mathsf{X}^{\ast}, ~~~
\mathsf{X}
:=
\mathsf{T_{12}} \mathsf{T_{22}}^{\,-1}.
\end{equation}
The overlap function $\braket*{ 0 }{ 1 }$ can be calculated by the Onishi formula \cite{OY66,BB69,RS80},
\begin{equation}\label{eq:onishi}
\braket*{ 0 }{ 1 }
=
\ev*{ e^{-i\hat{J}_y\beta} }{ \Phi }
=
\sqrt{\mbox{det} \mathsf{T_{22}} (\beta)}.
\end{equation}

\section{Proof for non-negativity of overlap function}\label{ap:nonneg}

We define a $A\times A$ square matrix $\mathsf{G}$ for a HF state $\ket{\Phi}$ 
whose elements are $\mel*{i'}{e^{-i\hat{J}_y \beta}}{i}$, 
where $i$ denotes the occupied s.p. state obtained by the HF calculation, 
\textit{i.e.}, $\ket{\Phi} = \prod_{i=1}^{A} a_i^\dagger \ket*{0}_c$.
%
We then have $\ev*{e^{ -i \hat{J}_y \beta}}{\Phi} = \det(\mathsf{G})$.
If $\hat{\mathcal{T}}\ket{\Phi} = \ket{\Phi}$
($\hat{\mathcal{T}}$ is the time-reversal operator),
$\ket{\Phi}$ contains time-reversal partners $\ket{i}$ and $\ket{\bar{i}}$,
which satisfy 
$\mel*{\bar{i'}}{e^{-i\hat{J}_y \beta}}{\bar{i}} 
= \mel*{i'}{e^{-i\hat{J}_y \beta}}{i}^{\ast}$
and 
$\mel*{\bar{i'}}{e^{-i\hat{J}_y \beta}}{i} 
= - \mel*{i'}{e^{-i\hat{J}_y \beta}}{\bar{i}}^{\ast}$
($\because~\hat{\mathcal{T}} e^{ -i \hat{J}_y \beta} \hat{\mathcal{T}}^{-1} 
= e^{ -i \hat{J}_y \beta}$).
Thus $\mathsf{G}$ has the following structure,
\begin{equation}\label{eq:T-part}
\mathsf{G}
= 
\mqty( \mathsf{A} & -\mathsf{B} \\ 
\,\mathsf{B}^{\ast} & ~\mathsf{A}^{\ast} )\,.
\end{equation}
It is proven that the structure of Eq. \eqref{eq:T-part} ensures 
$\det( \mathsf{G} ) \ge 0$ 
as follows.

The structure of Eq. \eqref{eq:T-part} derives the following property, 
\begin{equation}\label{eq:T-sym}
\mathsf{\Sigma}_y \mathsf{G} \mathsf{\Sigma}_y
= 
\mathsf{G}^{\ast} \, ,\quad
\mathsf{\Sigma}_y := \mqty( 0 & -i \\ i & ~0 ) \,.
\end{equation}
For the $\nu$-th eigenvalue and eigenvector of $\mathsf{G}$, 
\begin{equation}\label{eq:T-eigen}
\mathsf{G} \vb*{x}_\nu 
= 
\lambda_\nu \vb*{x}_\nu \,,
\end{equation}
there always exists a partner, 
\begin{equation}\label{eq:T-eigen2}
\mathsf{G} ( \mathsf{\Sigma}_y \vb*{x}_\nu^{\ast} )
= 
\lambda_\nu^{\ast} ( \mathsf{\Sigma}_y \vb*{x}_\nu^{\ast} ) \,,
\end{equation}
because of Eq. \eqref{eq:T-sym}.
Even when $\lambda_\nu = \lambda_\nu^{\ast}$, 
the eigenvectors are linearly independent because 
$( \mathsf{\Sigma}_y \vb*{x}_\nu^{\ast} )^\dagger \vb*{x}_\nu
= \vb*{x}_\nu^{\mathrm{T}} \mathsf{\Sigma}_y \vb*{x}_\nu = 0$.
While a matrix with the property \eqref{eq:T-sym} is not necessarily diagonalizable, 
the Jordan blocks associated by $\lambda_\nu$ and $\lambda_\nu^{\ast}$ have equal dimensions, 
as were given in the duality argument in Ref.~\cite{Na16}.
It is now proven, 
\begin{equation}
\det( \mathsf{G} ) 
= 
\prod_\nu \abs*{ \lambda_\nu }^2
\ge 0 \,.
\end{equation}
The non-negativity of matrices with the property \eqref{eq:T-part} was proven for quaternion matrices in Refs. \cite{Wi55,Zh97}.

\end{document}